\documentclass[prd,aps,showpacs,groupedaddress,eqsecnum,notitlepage,nofootinbib,superscriptaddress,10pt]{revtex4-2}
\usepackage{tikz}

\usepackage{mathtools}
\usepackage{bbm}
\usepackage{graphicx}
\usepackage{tikz,siunitx,mwe}
\usepackage{tensor}
\usepackage{epstopdf}
\usepackage{amsmath}
\usepackage{amsfonts}
\usepackage{slashed}
\usepackage{amssymb}
\usepackage{enumerate, color}
\usepackage{subfigure}
\usepackage{lipsum}
\usepackage{color}
\usepackage{graphicx,bm}
\usepackage[utf8]{inputenc}
\usepackage{eurosym}
\usepackage{scalerel}
\usepackage{float}
\usepackage{accents}
\usepackage[colorlinks=true,pdfstartview=FitV,linkcolor=blue,citecolor=blue,urlcolor=blue,breaklinks=true]{hyperref}
\usepackage{orcidlink}
\newcommand{\be}{\begin{equation}}
\newcommand{\ee}{\end{equation}}
\newcommand{\ben}{\begin{eqnarray}}
\newcommand{\een}{\end{eqnarray}}
\newcommand{\bes}{\begin{subequations}}
\newcommand{\ees}{\end{subequations}}
\usepackage{comment}
\def\bal#1\eal{\begin{align}#1\end{align}}

\def\bal#1\eal{\begin{align}#1\end{align}}
\begin{document}
\title{Effective Lifshitz-Born-Infeld black holes from general covariance breaking}
\author{D. C. Moreira\,\orcidlink{0000-0002-8799-3206}}
\email{moreira.dancesar@gmail.com}
\affiliation{Centro de Ciências, Tecnologia e Saúde, Universidade Estadual da Paraíba, 58233-000, Araruna, PB, Brazil}
\author{A. S. Lemos\,\orcidlink{0000-0002-3940-0779}}
\email{adiel@ufersa.edu.br}
\affiliation{Departamento de Ciências Exatas e Tecnologia da Informação, Universidade Federal Rural do Semi-Árido, 59515-000, Angicos, RN, Brazil}
\author{F. A. Brito\,\orcidlink{0000-0001-9465-6868}}
\email{fabrito@df.ufcg.edu.br}
\affiliation{Unidade Acadêmica de Física, Universidade Federal da Campina Grande, 58109-970, Campina Grande, PB, Brazil}
\affiliation{Unidade Acadêmica de Matemática, Universidade Federal de Campina Grande, 58429-970, Campina Grande, PB, Brazil}
\begin{abstract}
In this work we present an effective Lifshitz black hole solution with Born-Infeld electrodynamics and explore some of its properties. We discuss the mechanism for finding the solution, achieved through diffeomorphism invariance breaking, study the emergent causal structure, and analyze aspects of critical behavior and stability in the associated thermodynamics.
\end{abstract}

\maketitle
\section{Introduction}
Gauge/gravity duality has provided the theoretical physics community with a vast arsenal of techniques derived from string theory and gravitation for the study of strongly coupled systems in field theory and its applications \cite{ammon2015gauge,hartnoll2009lectures}. The idea of using gravity models with a specific group of isometries to study field theories located on the boundary of spacetime has been expanded and now offers to many fields of physics the opportunity to approach very difficult problems in a simplified way \cite{maldacena1999large,aharony2000large}. In this context, several types of geometries have been investigated to map different field models, and one of them has proven particularly useful for dealing with strongly coupled non-relativistic systems. Background geometries presenting non-relativistic scaling symmetry exhibit a scale anisotropy between space and time which allows to map strongly coupled non-relativistic field models, with applications in condensed matter systems. The simplest example of such geometries are Lifshitz spacetimes \cite{kachru2008gravity}, which are equipped with a dynamical exponent which measures the degree of deviation from the relativistic regime occurring in Anti-de Sitter (AdS) setups. As an example of applications of these geometries, we have Weyl semimetals, where electrons have a Fermi velocity much lower than the speed of light, and in the vicinity of the Weyl nodes, the emergent quasiparticles behave like massless fermions \cite{gursoy2013holographic, landsteiner2016quantum,landsteiner2020holographic,garcia2026zero}. Within this context, black holes play a prominent role, since their temperature induces a thermal bath on the dual field theory \cite{taylor2008non,taylor2016lifshitz,balasubramanian2009analytic,mann2009lifshitz,bertoldi2009black,danielsson2009black, cong2025holographic}, and therefore, its thermodynamical properties can be used to explore critical phenomena in quantum and condensed matter systems (see, for instance, Refs. \cite{deveciouglu2014lifshitz,brynjolfsson2013holographic,ayon2010analytic,ayon2009lifshitz,brito2020black,natsuume2018holographic,bazeia2015two,deveciouglu2011thermodynamics,gonzalez2011field,melnikov2019lifshitz,bravo2020thermodynamics,ayon2019microscopic,liu2014thermodynamics,tarrio2011black,tallarita2014holographic,li2014non,lu2014lifshitz,bravo2022lifshitz,moreira2024effective}). In particular, the study of charged Lifshitz black holes has made it possible, for example, to use gauge/gravity duality to analyze critical phenomena in superconductors \cite{brynjolfsson2010holographic}, complexity \cite{zhu2020holographic}, electrical conductivity \cite{jain2010universal}, and fermionic systems \cite{brynjolfsson2010black,fang2012holographic} in models with no conformal symmetry.

Running parallel to the topics presented above, works on classical field theories have emerged in the literature concerning extending Derrick's Theorem \cite{hobart1963instability,derrick1964comments} to classical probe scalar fields on curved backgrounds \cite{radmore1978non,palmer1979derrick,carloni2019derrick,mandal2021solitons}, as well as ways to circumvent them \cite{alestas2019evading,morris2021radially,mandal2021solitons,morris2022bps,moreira2022analytical,moreira2022erratum,moreira2023localized}. Some of the proposed evasions derive from Ref. \cite{bazeia2003new}, whose strategy consists of using explicitly coordinate-dependent potentials to construct kink-like solutions for scalar fields in flat spacetimes of any dimension. When these ideas are applied to static and radially symmetric spacetimes and also to Lifshitz spaces, it is possible to find well-behaved analytical field solutions using a first-order formalism captured through the Bogomol'nyi method \cite{morris2021radially,moreira2022analytical,moreira2022erratum,moreira2023localized,andrade2025analytical}. In these models, by setting aside the probe regime and exploring setups with gravity backreaction, it is possible to find effective solutions for neutral \cite{moreira2022scalar} and charged \cite{moreira2024charged} Lifshitz black holes by adjusting the terms responsible for the general covariance breaking with an auxiliary Maxwell field. The price to be paid when considering explicitly coordinate-dependent scalar potentials in systems with gravity backreaction is the breaking of diffeomorphism invariance, which presents some subtleties and inherently involves some difficulties \cite{bluhm2015explicit}. One of the major associated difficulties is the non-conservation of the energy-momentum tensor, which causes an incompatibility in Einstein's field equations since the Einstein tensor is always conserved due to the contracted Bianchi identity. Such a problem can be overcome by imposing by hand the divergence of the energy-momentum tensor to be zero on shell, which - despite introducing a constraint on the system - effectively restores the compatibility of Einstein's equations. Similar procedures are implemented, for example, in some cosmological scenarios where the Bianchi identity is applied on the energy-momentum tensor in models with non-dynamical background fields to constrain the field equations, making it possible to explore formal aspects, new explanations for dark matter, or models of emergent general relativity (See, for instance, Refs. \cite{bluhm2015spacetime,reyes2022cosmology,aydemir2025diffeomorphism} and related references.). For scenarios discussed in Refs. \cite{moreira2022scalar,moreira2024charged}, in addition to the terms violating general covariance, there is also an auxiliary Maxwell field which makes it possible to deal with the compatibility equation generated by the condition that the divergence of the energy-momentum tensor vanishes on shell. This idea of using an auxiliary Maxwell field in the search for analytical asymptotically Lifshitz black hole solutions is very useful and has already been investigated to, for example, study phase transitions in holographic superconductors \cite{brynjolfsson2010holographic} and find other charged solutions \cite{pang2010charged,dehghani2011charged,zangeneh2015thermodynamics}. A series of applications involving diffeomorphism invariance breaking mechanisms from the introduction of nondynamical background fields are addressed in \cite{anber2010breaking,cannone2015generalised,graef2015breaking,graef2017constraining,milgrom2019noncovariance,reyes2021hamiltonian, reyes2022modified,bluhm2021gravity,hidaka2015effective}. 

A natural extension of the study of charged black hole solutions involves considering nonlinear contributions from electrodynamics. The most famous prototype of these systems is the Born-Infeld model, which originally emerged from the interest in eliminating field divergences occurring in classical electromagnetism \cite{born1934foundations} and whose relevance was re-established after being observed in string theory scenarios \cite{gibbons2003aspects}. Among the wide range of models involving nonlinear electrodynamics, we can highlight that Born-Infeld black holes have been found and studied in both asymptotically flat \cite{garcia1984type,wiltshire1988black,de1994non,bronnikov2023regular} and non-zero cosmological constant spacetimes \cite{fernando2003charged,dey2004born,cai2004born}, exhibiting thermodynamical properties and causal structure quite distinct from its linear counterpart \cite{myung2008thermodynamics,banerjee2010note,banerjee2012critical,gunasekaran2012extended,dehyadegari2018reentrant,zou2014critical,jahromi2023nonlinear}. The thermodynamics and critical behavior of static Born-Infeld-AdS black holes in 3+1 dimensions, in particular, has a rich structure which, depending on the intensity of the non-linear corrections associated with Maxwell field dynamics, supports several distinct effects. Some of these effects occur in static solutions derived from linear electrodynamics, such as the Hawking-Page transition and Van der Waals-type phase transitions, which are usually associated with Schwarzschild-AdS and Reissner-Nordstrom-AdS solutions \cite{chamblin1999charged,chamblin1999holography,dolan2011cosmological,dolan2011pressure,kastor2009enthalpy,kubizvnak2012p,kubizvnak2017black}. Another interesting effect is the reentrant phase transition, which was originally found in rotating AdS solutions for $D\geq 6$ \cite{altamirano2013reentrant}, but which can also emerge in the Born-Infeld model for $D=4$ as a non-linear effect \cite{gunasekaran2012extended,dehyadegari2018reentrant}, not occurring for $D\geq 5$ \cite{zou2014critical}. See also \cite{naveena2021ruppeiner,bai2024reentrant,ali2025revisiting,guo2026complex,zhang2017reentrant,romero2024extended}.

In this work we are interested in obtaining and studying black hole solutions for $D\geq 4$ with nonlinear electrodynamics in asymptotically Lifshitz geometries arising from diffeomorphism invariance breaking, extending the solutions presented in Refs. \cite{moreira2022scalar, moreira2024charged} to setups with Born-Infeld dynamics. In particular, alongside the fields which break general covariance, we have two Maxwell fields with distinct roles. The first one supports the field nonlinearities associated with Born-Infeld coupling and has the function of providing electric charge to the background geometry. The second, in turn, acts as an auxiliary field whose function is to interact with the diffeomorphism invariance breaking terms, providing the system with new degrees of freedom which must be absorbed by the field equations after imposing the zero divergence condition on the energy-momentum tensor. Asymptotically Lifshitz black hole solutions with nonlinear electrodynamics appear in covariant systems with a Proca field and a massless Maxwell field coupled to an auxiliary antisymmetric tensor \cite{alvarez2014nonlinearly} and also in scenarios where the Maxwell field has Born-Infeld dynamics coupled to a dilaton \cite{zangeneh2017thermodynamics}. Furthermore, nonlinear electrodynamics on Lifshitz spacetimes are investigated in some holographic descriptions of superconductors involving probe fermion fields \cite{zangeneh2018optical,ghotbabadi2018one,naeimipour2021lifshitz}. Here, in addition we discuss the path to obtaining the solutions found, and also study its causal structure and thermodynamic, with critical behavior analyzed from the perspective of the extended phase space formalism, where the cosmological constant is used to define the black hole pressure and the mass of the black hole solution is associated to its enthalpy \cite{kastor2009enthalpy,dolan2011cosmological,dolan2011pressure,kubizvnak2012p}. In the scenario we are interested in, the breaking of diffeomorphism invariance has an effective character, and its existence, as a consequence of the field equations, depends on the presence of scaling anisotropy in the background geometry, since field solutions become trivial as we return to the relativistic regime, where the Born-Infeld-AdS solution emerges. As a consequence, ingredients usually found in Born-Infeld models, such as marginal mass, reentrant phase transitions, etc., have a direct dependence on the anisotropic scaling measure, whose values, for example, influence the causal structure, determining, along with the Born-Infeld parameter, the regions of the configuration space where solutions with a Cauchy horizon may or may not occur. This dependence particularly influences the thermal and critical behaviors, which exhibit distinct effects as the system moves away from the relativistic regime.

This work is organized as follows. In Sec. II, we present the model to be investigated, the field equations, and the compatibility equation derived from the conservation of the energy-momentum tensor. In Sec. III, we obtain the black hole solution and the field solutions derived from the field equations, in addition to discussing the causal structure. In Sec. IV, we study basic elements of the associated thermodynamics, and in Sec. V, we use these elements to address the critical behavior associated with the solution found, based on the extended phase space formalism. In Sec. VI, finally, we present our final considerations.

\section{Action and field equations}
Here we study an effective Einstein-Born-Infeld-scalar system whose action is given by
\begin{eqnarray}\label{action}
S=\int d^{D} x\sqrt{-g}\left(\frac{1}{2}R-\Lambda-\frac{1}{2}\nabla_a \phi\nabla^a\phi-V(x,\phi)-L\left(F\right)-\frac{1}{2}\varepsilon(x)\mathcal{F}_{ab}\mathcal{F}^{ab}\right),
\end{eqnarray}
where $g=\text{det}(g_{ab})$ is the determinant of the metric tensor, $R$ is  the curvature scalar, $\Lambda=-(D-2)(D+3z-4)/2\ell^2$ represents a negative cosmological constant and $\phi$ denotes a neutral scalar field whose behavior is driven by the explicitly coordinate-dependent scalar potential $V(x,\phi)$, where $x=\{x^a\}$, with $a=0,1,\cdots, D-1$. We also insert in the system two vector fields, $A_a$ and $B_a$. The first one is a vector field with Born-Infeld nonlinear dynamics given by
\begin{equation}\label{bidyn}
    L\left(F\right)=-2\beta^2\left(1-\sqrt{1+\frac{1}{2\beta^2}F_{ab}F^{ab}}\right), 
\end{equation}
where $\beta$ is the Born-Infeld parameter, $F=F_{ab}F^{ab}/2$ and $F_{ab}=\nabla_a A_b-\nabla_b A_a$. This vector field  approaches its usual dynamics for large $\beta$, since in this case one finds $L\left(F\right)\to\frac{1}{2}F_{ab}F^{ab}+\mathcal{O}\left(1/\beta^2\right)$ for $\beta\to\infty$. The field $B_a$, in turn, is an auxiliary ``Maxwell-like'' vector field with standard linear dynamics given in terms of the tensor $\mathcal{F}_{ab}=\nabla_a B_b-\nabla_b B_a$, which interacts with an effective magnetic permeability $\varepsilon(x)$. The field equations derived from the action \eqref{action} are
\begin{subequations}\label{ee}
\begin{eqnarray}
\Box \phi-\frac{\partial V}{\partial\phi}&=&0,\label{feq1}\\[1pt]
\nabla_a \left(\frac{ F^{ab}}{\sqrt{1+F/\beta^2}}\right)&=&0, \label{feq2}\\[3pt]
\nabla_a \left(\varepsilon(x) \mathcal{F}^{ab}\right)&=&0,\label{feq3}\\[3pt]
\mathcal{E}_{ab}=G_{ab}+\Lambda g_{ab}-T_{ab}&=&0,\label{feq4}
\end{eqnarray}
\end{subequations}
where $\Box=g^{ab}\nabla_a\nabla_b$ denotes the d'Alembertian operator, $G_{ab}=R_{ab}-g_{ab}R/2$ is the Einstein tensor, $T_{ab}$ represents the energy-momentum tensor, given by
\begin{eqnarray}\label{emt1}
\nonumber T_{ab}&=&\nabla_a\phi\nabla_b\phi-\frac{1}{2}g_{ab}\left(\nabla\phi\right)^2-g_{ab}V(x,\phi)+\frac{2F_{a}^{~c}F_{bc}}{\sqrt{1+F/\beta^2}}+2\beta^2g_{ab}\left(1-\sqrt{1+F/\beta^2}\right)+\\[8pt]
&~~~~&+\varepsilon(x)\left(2\mathcal{F}_{a}^{~c}\mathcal{F}_{bc}-\frac{1}{2}g_{ab}\mathcal{F}_{cd}\mathcal{F}^{cd}\right),~~~~~ 
\end{eqnarray}
 and $\mathcal{E}_{ab}$ denotes a ``zero tensor", defined for simplicity when dealing with field equations. 

By imposing explicit coordinate dependence on the system, we effectively introduce non-dynamical background fields into the model. It means that diffeomorphism invariance is not satisfied and consequently the energy-momentum tensor is no longer covariantly conserved, which implies that the right-hand side of Einstein's equation is no longer compatible with the Bianchi identity. This is a problem since the Einstein tensor always satisfies the contracted Bianchi identity, which has a geometric origin and comes from the symmetries of the Riemann tensor. One way to get around this problem and find new solutions is to look for effective configurations where dynamical and non-dynamical fields interact in such a way that $\nabla_a T^a_{~b}=0$, ensuring compatibility between both sides of Einstein's field equation. It leads us to the compatibility equation \cite{moreira2022scalar,moreira2024charged}
\begin{equation}\label{boundeq}
    \partial_a V(x, \phi)=-\frac{1}{2}\mathcal{F}_{bc}\mathcal{F}^{bc}\partial_a \varepsilon(x),
\end{equation}
which has to be considered side-by-side with the set of equations in Eq. \eqref{ee} and, in particular, assures that Eq. \eqref{feq4} is self-consistent. Equation \eqref{boundeq} acts as a constraint on the system and, in this way, it removes degrees of freedom from the field solutions. In particular, the presence of the auxiliary field $B_a$ in the constraint shows that its function, due to its ``Maxwell-like'' properties, is to provide the system with a conserved charge that mimics an electric charge and that must be absorbed from the final solution of the background geometry.
\section{Black hole solution}
We are interested in finding topological Lifshitz black hole solutions from the effective action \eqref{action} in background geometries with a generic structure given by
\begin{equation}\label{backmetric}
ds^2=-\left(\frac{r}{\ell}\right)^{2z}e^{2\nu(r)}dt^2+\left(\frac{\ell}{r}\right)^{2}\frac{dr^2}{e^{2\nu(r)}}+\left(\frac{r}{\ell}\right)^{2}\hat{\sigma}_{ij}(x^k)dx^i dx^j,
\end{equation}
where $\ell$ denotes a length scale, $z$ a dynamical exponent, $(x^0,x^1)=(t,r)$ are, respectively, time and radial coordinates and $i,j=2,\cdots, D-1$. The horizon metric $\hat{\sigma}_{ij}(x^k)$, with $2\leq k\leq D-1$, describes a closed $(D-2)$-dimensional Einstein manifold $\hat{\Sigma}_{\gamma}$ whose  Ricci tensor is $\hat{R}_{ij}=(D-3)\gamma\hat{\sigma}_{ij}/\ell^2$ where $\gamma=0,\pm1$, thus holding spherical $(\gamma=1)$, planar $(\gamma=0)$ or hyperbolic $(\gamma=-1)$ topologies. For $z\to 1$ we retrieve AdS$_{D}$ setups and the standard Lifshitz spacetime \cite{kachru2008gravity,taylor2016lifshitz} is found in the limit $\nu(r)\to 0$ for $\hat{\sigma}_{ij}=\delta_{ij}$.
We also assume that all fields, magnetic permeability and scalar potential only have radial dependence, i.e.,
\begin{eqnarray}
\phi=\phi(r), ~A=A(r)dt,~B=B(r)dt,~\varepsilon(x)=\varepsilon(r)~\text{and}~V(x,\phi)=V(r,\phi(r)).
\end{eqnarray}
In this way, Eq. \eqref{feq2} and Eq. \eqref{feq3} becomes
\begin{subequations}\label{vf}
    \begin{eqnarray}
         A'(r)&=&\frac{q/\ell}{\sqrt{1+\frac{q^2/\ell^2}{\beta^2}\left(\frac{\ell}{r}\right)^{2\left(D-2\right)}}}\left(\frac{\ell}{r}\right)^{D-z-1},\label{eqvec1}\\[4pt]
         B'(r)&=&\frac{\widetilde{q}/\ell}{\varepsilon(r)}\left(\frac{\ell}{r}\right)^{D-z-1},\label{eqvec2}
    \end{eqnarray}
\end{subequations}
where the prime denotes derivation with respect to the radial coordinate and $\left(q,\widetilde{q}\right)$ are integration constants related to the conserved charges associated to the Maxwell field $A_a$ and the auxiliary field $B_a$, expressed as
\begin{subequations}
\begin{eqnarray}
Q&=&-\frac{1}{4\pi}\oint_{\partial\Sigma} d^{D-2}x \sqrt{|h^{(2)}|}\frac{n_a s_b F^{ab}}{\sqrt{1+F/\beta^2}}=\frac{\omega_{D-2}^{(\gamma)}}{4\pi\ell}q,\\[3pt]
\widetilde{Q}&=&-\frac{1}{4\pi}\oint_{\partial\Sigma} d^{D-2}x \sqrt{|h^{(2)}|} n_a s_b \varepsilon (x)\mathcal{F}^{ab}=\frac{\omega_{D-2}^{(\gamma)}}{4\pi\ell}\widetilde{q},~~~~~~~~
\end{eqnarray}
\end{subequations}
respectively. In addition, $n_a$ and $s_a$ are timelike and spacelike unit normal vectors to the  surface $\partial\Sigma$ defined at fixed $(r,t)$, equipped with the induced metric $h_{ij}^{(2)}=\left(\frac{r}{\ell}\right)^2\hat{\sigma}_{ij}$, such that $h^{(2)}=\text{det}\left(h^{(2)}_{ij}\right)$, and $\omega_{D-2}^{(\gamma)}=\oint_{\Hat{\Sigma}_\gamma}d^{D-2}x\sqrt{|\Hat{\sigma}_\gamma|}$ denotes the volume of $\hat{\Sigma}_{\gamma}$.

In this setup, the $(r,r)$ and $(t,t)$ components of $\mathcal{E}^a_{~b}=0$ in Eq. \eqref{feq4} are
\begin{subequations}\label{eeq}
\begin{eqnarray}
\nonumber\label{err} \mathcal{E}^{r}_{~r}&=&\frac{(D-2)/2\ell^2}{r^{D+2z-4}}\left(r^{D+2z-3}e^{2\nu}\right)'-\frac{1}{2}\left(\frac{r}{\ell}\right)^{2}\phi'^2 e^{2\nu}+V(r,\phi)+\Lambda-\frac{\hat{\gamma}}{2r^2}-2\beta^2\left(1-\sqrt{1+\frac{q^2/\ell^2}{\beta^2}\left(\frac{\ell}{r}\right)^{2(D-2)}}\right)+\\[8pt]
&&~~~~~~~~~~+\frac{\widetilde{q}^2/\ell^2}{\varepsilon(r)}\left(\frac{\ell}{r}\right)^{2(D-2)},\\[8pt]
\nonumber \label{ett}\mathcal{E}^{t}_{~t}&=&\frac{(D-2)/2\ell^2}{r^{D-2}}\left(r^{D-1}e^{2\nu}\right)'+\frac{1}{2}\left(\frac{r}{\ell}\right)^{2}\phi'^2 e^{2\nu}+V(r,\phi)+\Lambda-\frac{\hat{\gamma}}{2r^2}-2\beta^2\left(1-\sqrt{1+\frac{q^2/\ell^2}{\beta^2}\left(\frac{\ell}{r}\right)^{2(D-2)}}\right)+\\[8pt]
&&~~~~~~~~~~+\frac{\widetilde{q}^2/\ell^2}{\varepsilon(r)}\left(\frac{\ell}{r}\right)^{2(D-2)},
\end{eqnarray}
\end{subequations}
where $\hat{\gamma}=(D-2)(D-3)\gamma$ and the remaining equations $\mathcal{E}^i_{~j}=0$ are left for checking. One finds from the difference $\mathcal{E}^{r}_{~r}-\mathcal{E}^{t}_{~t}=0$ that the scalar field must satisfy the first-order differential equation
\begin{equation}\label{phieq}
    \left(\frac{d\phi}{dr}\right)^2=\frac{(D-2)(z-1)}{r^2},
\end{equation}
which is valid for $z\geq1$ and whose solution is given  by
\begin{eqnarray}\label{sfs}  
\phi(r)=\phi_0\pm\sqrt{(D-2)(z-1)}\ln(r/\ell),
    \end{eqnarray}
where $\phi_0$ denotes an integration constant. Moreover, the analytical solution for the Maxwell field can be found directly by integrating Eq. \eqref{eqvec1}, leading to
\begin{equation}\label{maxwellsol}
A(r)=\Phi-\frac{q}{D-z-2}\left(\frac{\ell}{r}\right)^{D-z-2}\,_2F_1\left(\frac{1}{2},\frac{D-z-2}{2(D-2)};\frac{3D-z-6}{2(D-2)};-\frac{q^2/\ell^2}{\beta^2}\left(\frac{\ell}{r}\right)^{2(D-2)}\right),
\end{equation}
which diverges for $z=D-2$ in cases where $q\neq0$ and where $\Phi$ denotes an integration constant which, as shown below, is related to the electrostatic potential calculated at the event horizon of the black hole. The solution for the auxiliary field $B(r)$, in turn, depends on the expression to be found for the magnetic permeability, associated with the compatibility equation in \eqref{boundeq}.

In order to solve the equations \eqref{err} and \eqref{ett}, we have to make an appropriate choice for the potential $V(r,\phi)$, consistent with Einstein's equations \eqref{feq4} and the scalar field equation \eqref{feq1}, in addition to providing a route to capture an analytical expression for magnetic permeability through the compatibility equation \eqref{boundeq}. All these requirements are satisfied if we take the following {\it ansatz} for the scalar potential \cite{moreira2022scalar,moreira2024charged}:
\begin{equation}\label{1potmodel}
V(r,\phi)=\frac{1}{2}e^{-2\nu}\left(\frac{\ell}{r}\right)^{2(D+z-2)}\left(\frac{dW}{d\phi}\right)^{2}+U(r),
\end{equation}
where $dW/d\phi$ and $U(r)$ are auxiliary functions to be specified from the field equations. The first term of the scalar potential \eqref{1potmodel} describes the self-interaction of the scalar field coupled to an explicitly coordinate-dependent factor which breaks diffeomorphism invariance. It allows us to perform an order reduction in the scalar field equation \eqref{feq1}, which becomes
\begin{equation}\label{1ordereq}
\frac{d\phi}{dr}=\pm  \left(\frac{\ell}{r}\right)^{D+z-1}\frac{dW}{d\phi}e^{-2\nu},
\end{equation}
and, since the field solution \eqref{sfs} is invertible, one can find an analytical expression for $dW/d\phi$ from Eq. \eqref{1ordereq}. In probe regime, this auxiliary function is responsible for modeling classical spatially localized scalars with minimal energy on curved, static backgrounds \cite{moreira2022scalar}. The second term of the potential is constituted by a scalar function $U(r)$, which acts as a nondynamical degree of freedom of the system and, despite not interacting with the scalar and vector fields, has a role in determining the magnetic permeability. By handling  Eq. \eqref{phieq}, Eq. \eqref{1potmodel} and Eq. \eqref{1ordereq}, one can show that the  compatibility equation \eqref{boundeq} is satisfied by the pair of functions 
\begin{subequations}
\begin{eqnarray}
\label{dielfunc}   \frac{1}{\varepsilon(r)}&=&\frac{\left(z-1\right)\left(D+z-2\right)}{2\widetilde{q}^2}\left(\frac{r}{\ell}\right)^{2(D-2)}e^{2\nu},~~~~\\[2pt]
 \label{ur}   U(r)&=&-\frac{z\left(z-1\right)}{2\ell^2}e^{2\nu}.\label{ufunc}
\end{eqnarray}
\end{subequations}
For $z\to 1$, the solutions above go to zero and the scalar field \eqref{sfs} becomes trivial, revealing that the effective solution we are looking for must exactly retrieve the Born-Infeld AdS black hole in this limit \cite{gunasekaran2012extended,zou2014critical}. In particular, Eq. \eqref{dielfunc} can be rewritten as 
\begin{equation}
    \frac{\widetilde{q}^2/\ell^2}{\varepsilon(r)}\left(\frac{\ell}{r}\right)^{2(D-2)}=\frac{(z-1)(D+z-2)}{2\ell^2}e^{2\nu}
\end{equation}
and considered in combination with the relations \eqref{phieq}, \eqref{1potmodel}, \eqref{1ordereq} and \eqref{ufunc} to manipulate and express the equations \eqref{err} and \eqref{ett} as
\begin{eqnarray}\label{err=ett}
    \mathcal{E}^r_{~r}=\mathcal{E}^t_{~t}=\frac{(D-2)/2\ell^2}{r^{D+3z-5}}\left(r^{D+3z-4}e^{2\nu}\right)'+\Lambda-\frac{\hat{\gamma}}{2r^2}-2\beta^2\left(1-\sqrt{1+\frac{q^2/\ell^2}{\beta^2}\left(\frac{\ell}{r}\right)^{2(D-2)}}\right)=0,
\end{eqnarray}
which finally provides the solution for the horizon function,
\begin{eqnarray}\label{metricsol}
\nonumber e^{2\nu(r)}&=&1-2m\left(\frac{\ell}{r}\right)^{\Delta}+\frac{\gamma_z}{r^2}+\frac{4\ell^2\beta^2}{(D-2)\Delta}\left(1-\sqrt{1+\frac{q^2/\ell^2}{\beta^2}\left(\frac{\ell}{r}\right)^{2(D-2)}}\right)+\\[8pt]
&~~~~&+\frac{4q^2}{(D-3z)\Delta}\left(\frac{\ell}{r}\right)^{2(D-2)}\,_2F_1\left(\frac{1}{2},\frac{D-3z}{2(D-2)}~;1+\frac{D-3z}{2(D-2)};-\frac{q^2/\ell^2}{\beta^2}\left(\frac{\ell}{r}\right)^{2(D-2)}\right),
\end{eqnarray}
where, for simplicity, we use $\gamma_z=(D-3)\gamma\ell^2/(D+3z-6)$ and $\Delta=D+3z-4$. The solution \eqref{metricsol} is well-behaved for $D+3z-6\neq 0$ and for $D\neq 3z-2(n+1)(D-2)$, with $n=-1,0,1,2,\cdots$. It retrieves Lifshitz spacetime in the limit $r\to\infty$ since we always have $\Delta>0$, so that $e^{2\nu(r)}\to 1$ for any $\beta\neq0$. Note that for $z\to 1$ we retrieve the Born-Infeld black hole solution addressed in \cite{dey2004born,cai2004born,banerjee2012critical,gunasekaran2012extended, zou2014critical}, while the limit $\beta\to\infty$ exactly reproduce the effective Lifshitz black hole solution recently found in \cite{moreira2024charged}. 

\begin{figure}[t]
  \centering
  \begin{tabular}{ c @{\quad} c }
  \includegraphics[scale=0.4]{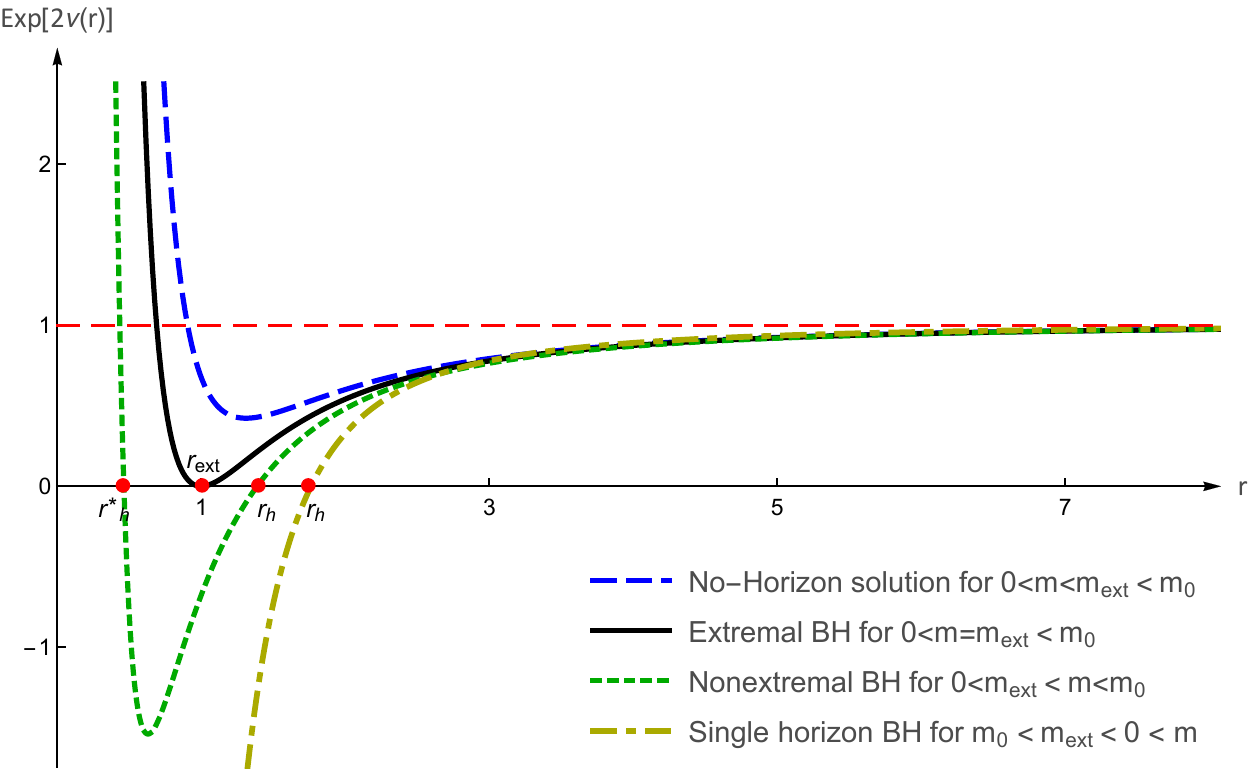} &
    \includegraphics[scale=0.4]{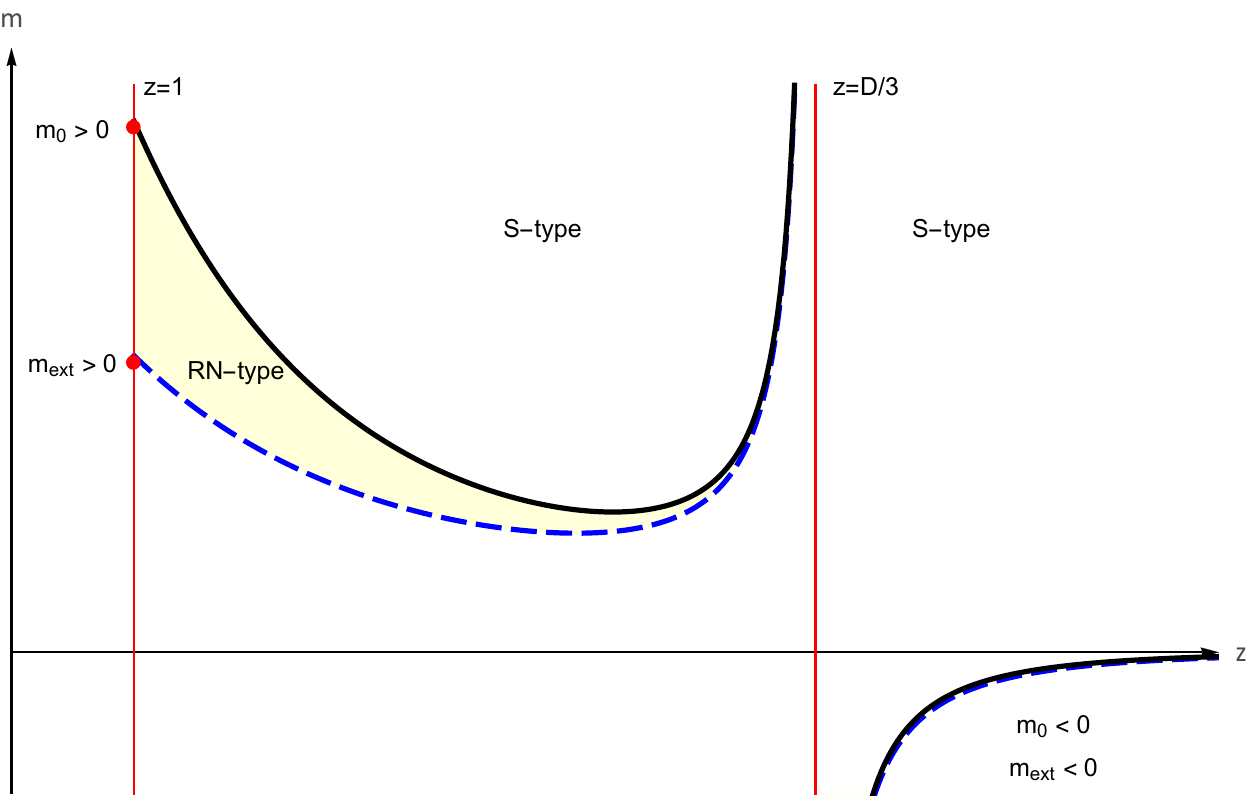}  \\
    \small (a) &
      \small (b)
  \end{tabular}
  \vspace*{8pt}
  \caption{Causal structure of the solution \eqref{metricsol} for $0< m\neq m_0$. In (a), we use $\beta=0.1$, $D=4$, $\gamma=1$, $r_{ext}=1$, $q=q_{ext}$, $\ell=3/2$ and define an auxiliary quantity $\delta m=(m_0-m_{ext})/2$, for direct comparison with the marginal mass. For $1<z<D/3$ (RN-type scenarios), we use $z=(1+\frac{D}{3})/2$ and find no horizon solutions for $m=m_{ext}-|\delta m|$ (blue, dashed), extremal black holes for $m=m_{ext}$ (black, solid) and non-extremal black holes for $m=m_{ext}+|\delta m|$ (green, dotted). For $z>D/3$ (S-type scenarios) we use $z=0.25+D/3$ and $m=m_{ext}+20|\delta m|$ in order to have $m>0$ since in this case one finds $m_{ext}<0$. In (b) we show schematically how the suppression of the configuration space of solutions with Cauchy horizons  (green region) occurs. Changing the topology does not qualitatively affect the results illustrated here.}\label{fig1} 
\end{figure}

In the asymptotic regime, the solution \eqref{metricsol} behaves as 
\begin{equation}
    e^{2\nu}\approx 1-2m\left(\frac{\ell}{r}\right)^{\Delta}+\frac{\gamma_z}{r^2}+q_z^2\left(\frac{\ell}{r}\right)^{2(D-2)}-\frac{1}{(D-2)(3D-3z-4)}\frac{q^4/\ell^2}{2\beta^2}\left(\frac{\ell}{r}\right)^{4(D-2)}+\cdots ,
\end{equation}
with $q_z^2=2q^2/\big((D-2)(D-3z)\big)$. It shows that nonlinear terms from electrodynamics are suppressed for large $r$, since far from the origin we have only small deviations from the linear case investigated in \cite{moreira2024charged}. These nonlinearities, however, strongly affect the causal structure of the solution found, since in the region $r\approx0$ we have
\begin{eqnarray}\label{solhorizon}
e^{2\nu(r)}&\approx&2\left(m_0\!-\!m\right)\!\left(\frac{\ell}{r}\right)^{\!\Delta}\!\!+\!\frac{\gamma_z}{r^{2}}-\frac{4\beta\ell q}{(3z\!-\!2)(D\!-\!2)}\!\left(\frac{\ell}{r}\right)^{\!D-2}\!\!+\frac{2q^2}{\Delta(\Delta\!+\!D\!-\!2)}\!\left(\frac{\beta\ell}{q}\right)^{\!3}\!\left(\frac{r}{\ell}\right)^{D-2}\!\!+\mathcal{O}\left(\left(\frac{r}{\ell}\right)^{3(D-2)}\right),~~~~
\end{eqnarray}
where a charge-induced {\it marginal mass}, given by 
\begin{equation}\label{marginalmass}
    m_0=\frac{\Gamma\left(\frac{3z-2}{2(D-2)}\right)\Gamma\left(\frac{D-3z}{2(D-2)}\right)\left(\beta\ell\right)^{\frac{D-3z}{D-2}}}{(D-2)\sqrt{\pi}}\frac{q^{\Delta/(D-2)}}{\Delta},
\end{equation}
emerges, generalizing results found in \cite{dey2004born,banerjee2012critical,zou2014critical}. For fixed $\beta$, $m_0$ goes to zero in the limit $q\to 0$  and is well behaved for all allowed values of $(D, z)$ from the solution \eqref{metricsol}, being negative for $D-3z<0$ and positive for $D-3z>0$. In particular, for $D-3z<0$ we have $m_0\to(0,-\infty)$ when $\beta\to(\infty,0)$ and for $D-3z>0$ one finds $m_0\to(0,\infty)$ when $\beta\to(0,\infty)$. Neither the marginal mass nor the background solution found are defined for $D=3z$. In any dimension, if $m\neq m_0$, the region $r\approx 0$ is dominated by the ``mass parameter'' contribution in Eq. \eqref{solhorizon} and all the ingredients of the emerging causal structure can be explained by analyzing the horizon function \eqref{metricsol}. Indeed, if $m_0<m$ we have $e^{2\nu(r\to 0)}\to -\infty$, which indicates the existence of a single (event) horizon located at some $r=r_h$. In this case this kind of behavior has been called Schwarzschild-type (S-type) in the literature. For $m_0>m$ we have $e^{2\nu(r\to 0)}\to \infty$, indicating scenarios with zero, one or two horizons. This kind of behavior has been called Reissner-Nordstr\"om-type (RN-type). The case where $m=m_0$ has very distinct properties and is detailed below. Here we only consider scenarios where $m>0$ and therefore setups with $m_0<0$, where $z>D/3$, can only provide S-type solutions. In this way, we only have RN-type solutions in setups with $1<z<D/3$. 

By analyzing the extreme values of $e^{2\nu(r)}$ for $m_0> m$, it is possible to identify which conditions must be satisfied for RN-type solutions to exhibit horizons, as illustrated in Fig. (\ref{fig1}a). This approach reveals that the solution \eqref{metricsol} must have at least one horizon if there is an absolute minimum for a given radius $r=r_0$ such that the inequality
\begin{eqnarray}\label{ineq}
 \nonumber  1+(\Delta-2)m\left(\frac{\ell}{r_0}\right)^{\Delta}&\leq& \frac{\Delta-2}{\Delta}\frac{2q^2}{D-3z}\left(\frac{\ell}{r_0}\right)^{2(D-2)}\,_2F_1\left(\frac{1}{2},\frac{D-3z}{2(D-2)}~;1+\frac{D-3z}{2(D-2)};-\frac{q^2/\ell^2}{\beta^2}\left(\frac{\ell}{r_0}\right)^{\!\!2(D-2)}\right)-\\[5pt]
 &-&\frac{4\ell^2\beta^2}{(D-2)\Delta}\left(1-\sqrt{1+\frac{q^2/\ell^2}{\beta^2}\left(\frac{\ell}{r_0}\right)^{2(D-2)}}\right),
\end{eqnarray}
is satisfied. This bound is saturated in the extremal limit, with extremal charge and mass parameters given by
\begin{subequations}
\begin{eqnarray}
q^2_{ext}&=&\frac{\ell^2}{2}\left(\frac{r_{ext}}{\ell}\right)^{2(D-2)}\left(\frac{\hat{\gamma}}{r^2_{ext}}+\frac{(D-2)\Delta}{\ell^2}\right)\left(1+\frac{1}{8\beta^2}\left(\frac{\hat{\gamma}}{r^2_{ext}}+\frac{(D-2)\Delta}{\ell^2}\right)\right),~\\[5pt]
m_{ext}&=&\frac{1}{\Delta}\!\left(\frac{r_{ext}}{\ell}\right)^{\!\Delta}\!\!\left(\frac{\gamma_z}{r^2_{ext}}\!+\!\frac{2q^2_{ext}}{D-3z}\left(\frac{\ell}{r_{ext}}\right)^{\!2(D-2)}\!\!\!\,_2F_1\!\left(\frac{1}{2},\frac{D-3z}{2(D-2)};1\!+\!\frac{D-3z}{2(D-2)};-\frac{q^2_{ext}}{\ell^2\beta^2}\left(\!\frac{\ell}{r_{ext}}\right)^{\!2(D-2)}\right)\!\right),~~~
\end{eqnarray}
\end{subequations}
respectively, where $r=r_{ext}$ represents the extremal radius and for simplicity we choose to leave $m_{ext}$  written in terms of $q_{ext}^2$. In this way, RN-type black hole solutions occur for $m\geq m_{ext}>0$ and, consequently, due to the competition between the marginal and extremal masses, one must find such solutions with distinct horizons (an event horizon and a Cauchy horizon) only in a mass range defined by $m_0>m> m_{ext}$. For $m=m_{ext}$, in particular, the horizons become degenerate. Both the marginal and extremal masses diverge in the limit $\beta\to 0$ for $1<z<D/3$, indicating that the more intense the nonlinearity of the electromagnetic field is, the more difficult is the formation of two-horizon configurations. Furthermore, as illustrated in Fig. (\ref{fig1}b), for each fixed $\beta$, increasing the scaling anisotropy restricts the mass range accessible to RN-type solutions, making the emergence of these solutions increasingly difficult as $z\to D/3$, since in this limit both masses increase sharply as their values approach each other, making the very narrow mass range difficult to reach. Like the marginal mass, the extremal mass is not defined for $D=3z$, being positive (negative) for $D>3z~(D<3z)$. It is also worth mentioning that, in the same way as in other Born-Infeld AdS black hole solutions \cite{gunasekaran2012extended,zou2014critical}, the case where $D=4$ is special, since in these setups, in addition to the conditions addressed so far, we still have a constraint given by the temperature behavior, which must be zero in the extremal limit. Taking this constraint into account, we observe that solutions with two horizons can only occur if $\beta q>\gamma/(2\ell)$, as we discuss in the next section. For $D \geq 5$, RN-type solutions exist for any $\beta q>0$.

\begin{figure}[t]
  \centering
  \begin{tabular}{c@{\quad}c@{\quad}c}
  \includegraphics[scale=0.2675]{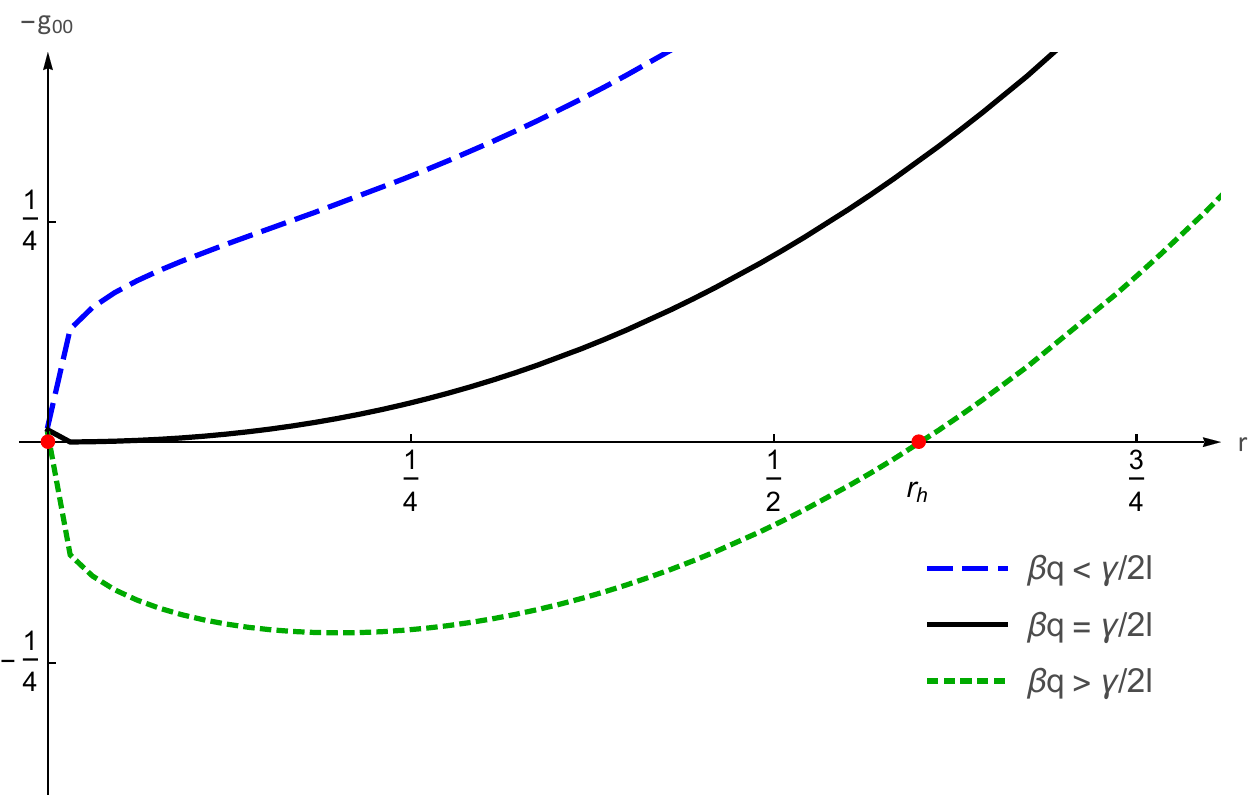} &
  \includegraphics[scale=0.2675]{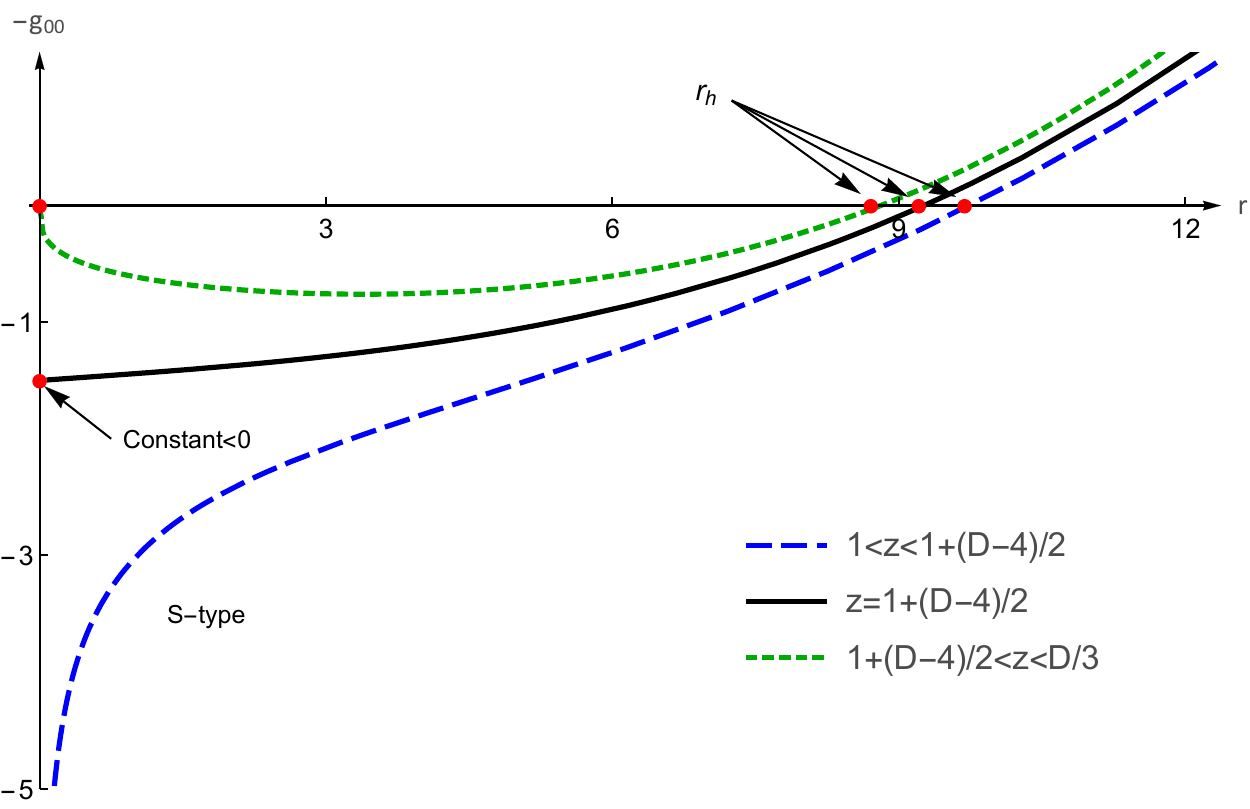} & 
  \includegraphics[scale=0.2675]{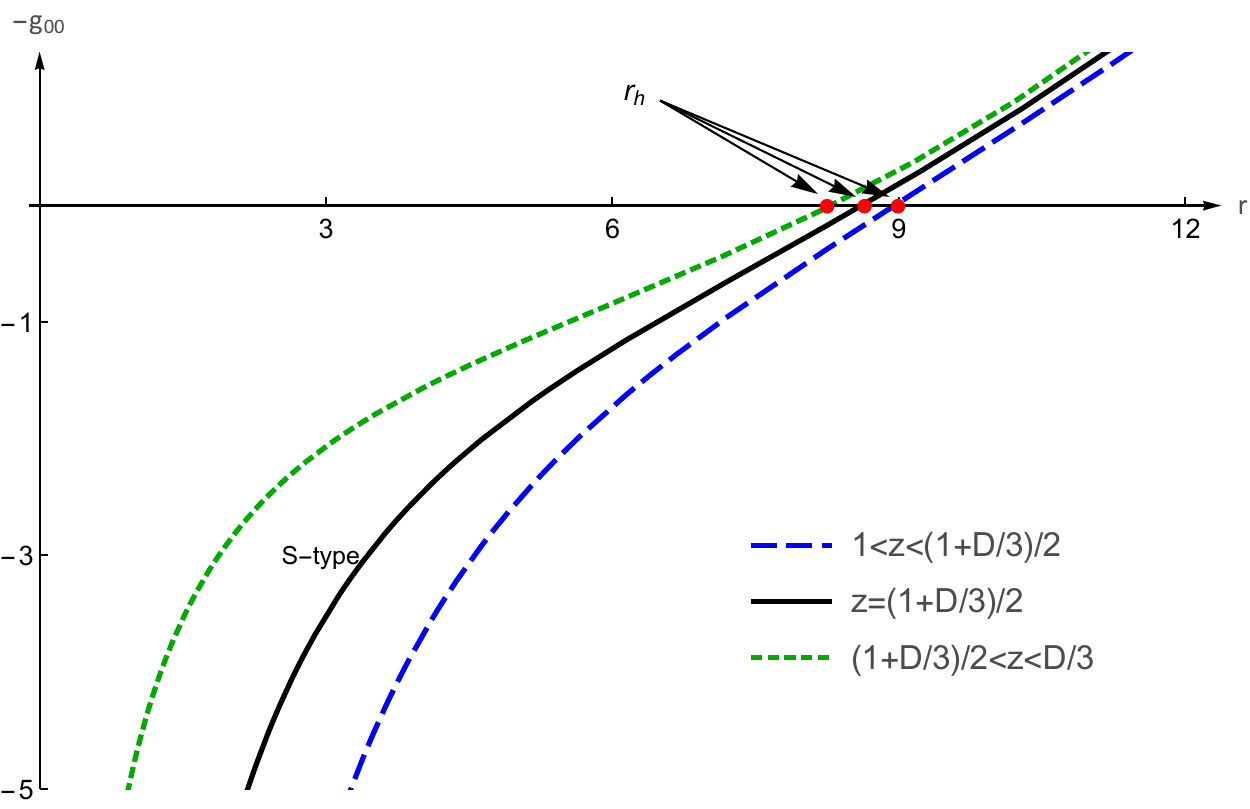}\\
    \small (a) &
    \small (b) &
    \small (c)
  \end{tabular}
  \vspace*{8pt}
  \caption{Causal structure for $1<z<D/3$ and $m=m_0$, with $\beta=1/10$ and $\gamma=1$. In (a) we consider $D=4$ and different $\beta q$-products, with $\ell=1$, $z=9/8$ and $\beta q=(1/2\ell)/2~\text{(blue, dashed)},~ 1/2\ell~\text{(black)},~ 3(1/2\ell)/2~\text{(green, dotted)}$. In (b) we illustrate the scenario for $D=5$ with different values of $z$, with $\ell=10$, $q=45/16$ and $z= ((D-2)/2)-0.125~\text{(blue, dashed)},~(D-2)/2~\text{(black)}, (D-2)/2+0.15~\text{(green, dotted)}$. In (c) we have $D=6$ and different values of $z$, with $\ell=10$, $q=45/16$ and $z= (1+D/3)/2-0.125~\text{(blue, dashed)},~(1+D/3)/2~\text{(black)}, (1+D/3)/2+0.15~\text{(green, dotted)}$.}\label{fig2}
\end{figure}

For $m=m_0$, the causal structure drastically changes and cannot be determined by the horizon function alone, but rather by the full metric component $g_{00}=-\left(\frac{r}{\ell}\right)^{2z}e^{2\nu(r)}$. In this case, in the region $r\approx 0$ we have
\begin{eqnarray}\label{intpot} {\large
  \Bigg.-g_{00}\Bigg|_{r\approx 0,~m  = m_0}\approx\left\{
\begin{array}{l}
\frac{2\ell}{3z-2}\left(\frac{\gamma}{2\ell}- \beta q\right)\left(\frac{r}{\ell}\right)^{2(z-1)}+\cdots, ~~\text{for}~~ D=4,\\[10pt]
-\frac{4\beta \ell q}{(3z-2)(D-2)}\left(\frac{\ell}{r}\right)^{D-2z-2}+\cdots, ~~\text{for}~~ D>4. 
\end{array}
\right.}
\end{eqnarray}
For $z\to 1$, we retrieve the results discussed in \cite{fernando2003charged,gunasekaran2012extended}, 
but for $z>1$ we have substantial differences due to the presence of anisotropic scaling. Indeed, for $D=4$ and any topology,  if $z>1$ we observe $-g_{00}\to 0$ as $r\to 0$, which indicates that the background metric has a singularity at $r=0$ for any allowed value of the dynamical exponent, as illustrated in FIG. (\ref{fig2}a). The concavity analysis of the curve reveals that  such a singularity is naked for $\beta q\leq \gamma/(2\ell)$ and dressed by a single event horizon if $\beta q>\gamma/(2\ell)$. Note that in order to ensure $m_0\in \mathbb{R}$ it is necessary to have $q>0$, which implies that these distinct qualitative behaviors only occur in setups with spherical topology $(\gamma=1)$, since if $\gamma=-1, 0$ we have $\gamma/(2\ell)\leq0$ and, consequently, in these cases we have single-horizon solutions for any $q>0$. On the other hand, scenarios with $D>4$ present single-horizon black holes whose behavior in the limit $r\to 0$ is determined by a competition between the spacetime dimension and the critical exponent. In particular, if $D-2z-2>0$ we have $1<z<\text{min}\{1+(D-4)/2, D/3\}$ and $-g_{00}\to-\infty$, which leads us to S-type solutions. If $D-2z-2=0$, we have $1<z=1+(D-4)/2<D/3$ and $-g_{00}$ approaches a negative - finite - value. Finally, if $D-2z-2<0$, we have $1+(D-4)/2<z<D/3$ and $-g_{00}\to 0$, where a dressed singularity arises. Note that the last two cases only occur if $1+(D-4)/2<D/3$, which only holds true for $D=5$, whose case is depicted in FIG. (\ref{fig2}b). Therefore, setups with $m=m_0$ and $D\geq 6$ only allow the emergence of S-type solutions, as shown in FIG. (\ref{fig2}c).

\section{Thermodynamics and Smarr-type relation}

\begin{figure}[t!]
  \centering
  \begin{tabular}{ c @{\quad} c @{\quad} c }
        \includegraphics[scale=0.275]{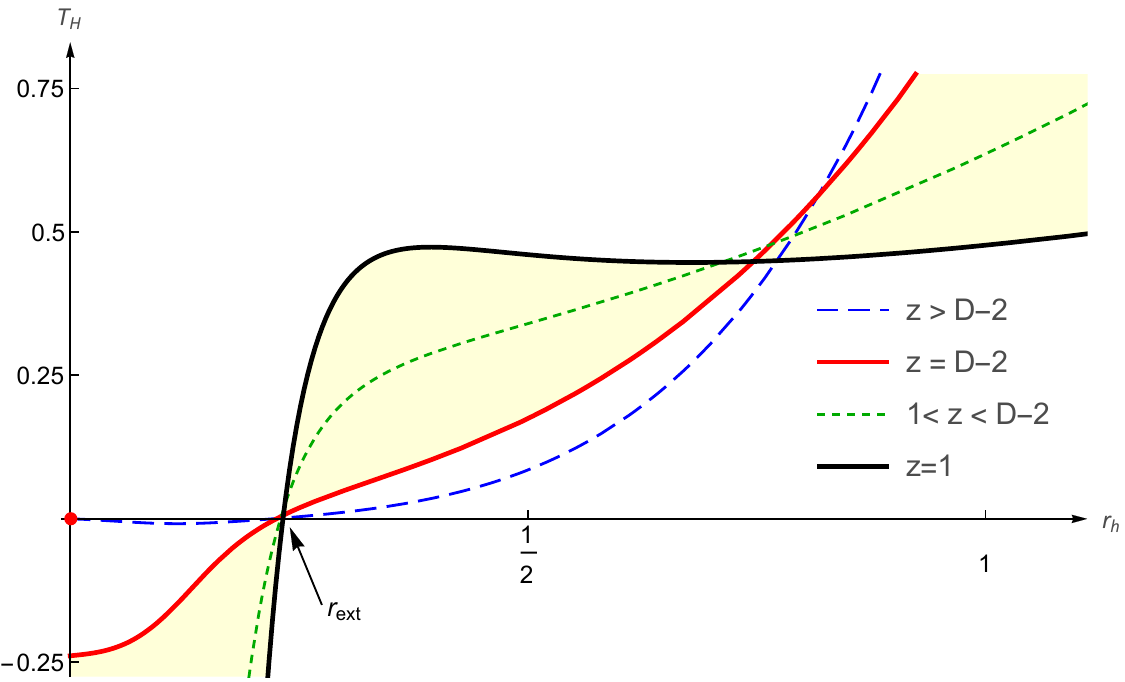}&
  \includegraphics[scale=0.2675]{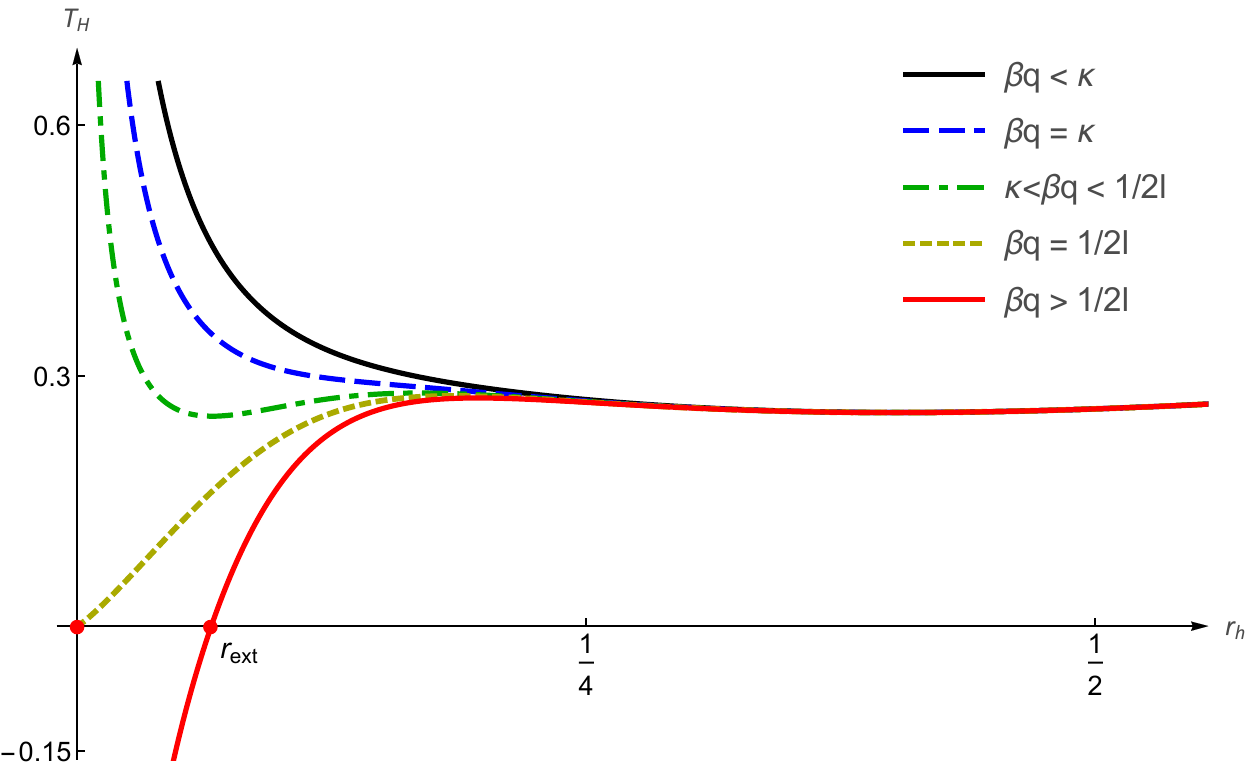}&
  \includegraphics[scale=0.2675]{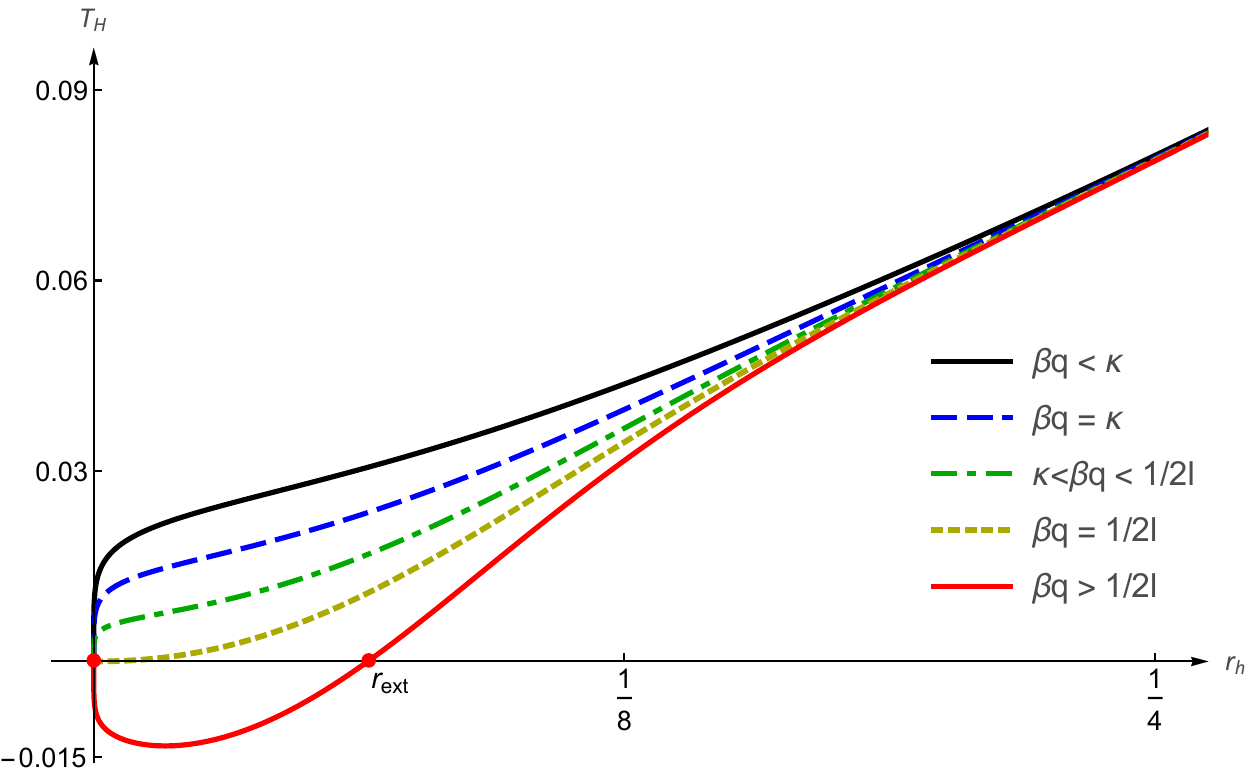} 
   \\
    \small (a)  &
      \small (b) &
      \small (c) \\
  \includegraphics[scale=0.2675]{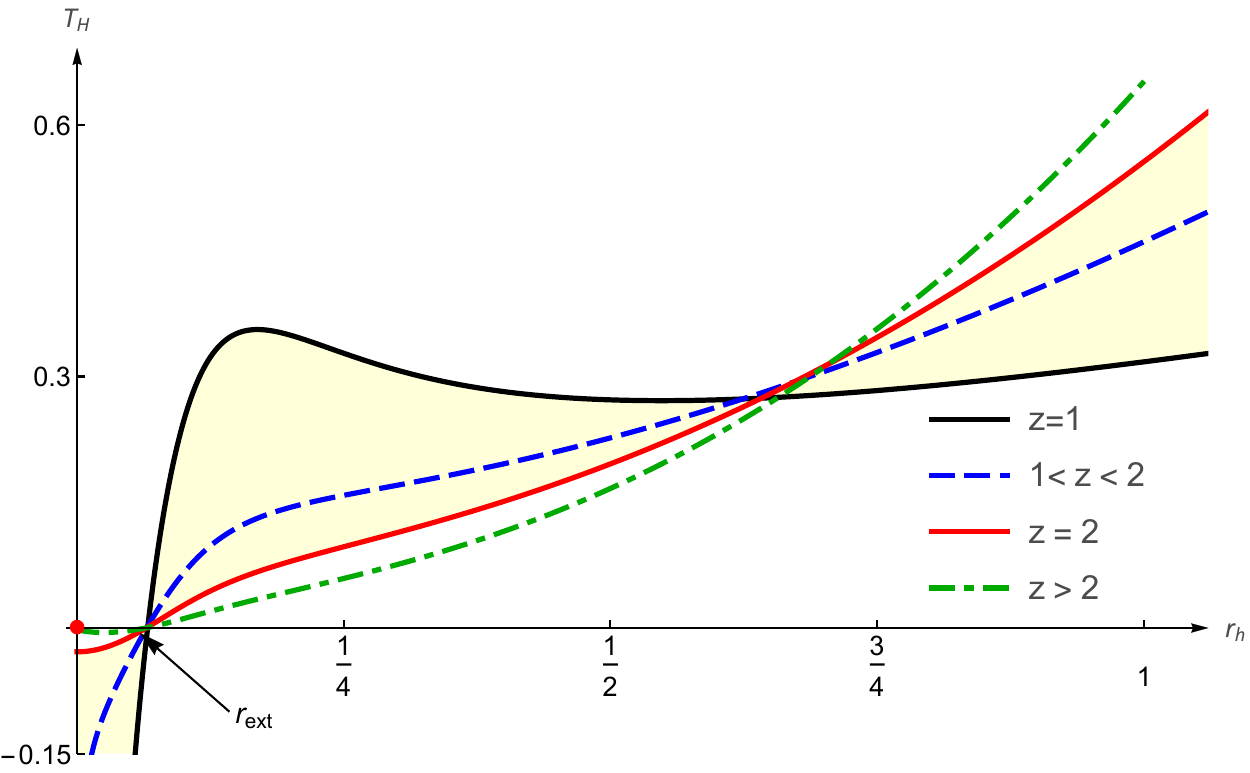} &   \includegraphics[scale=0.2675]{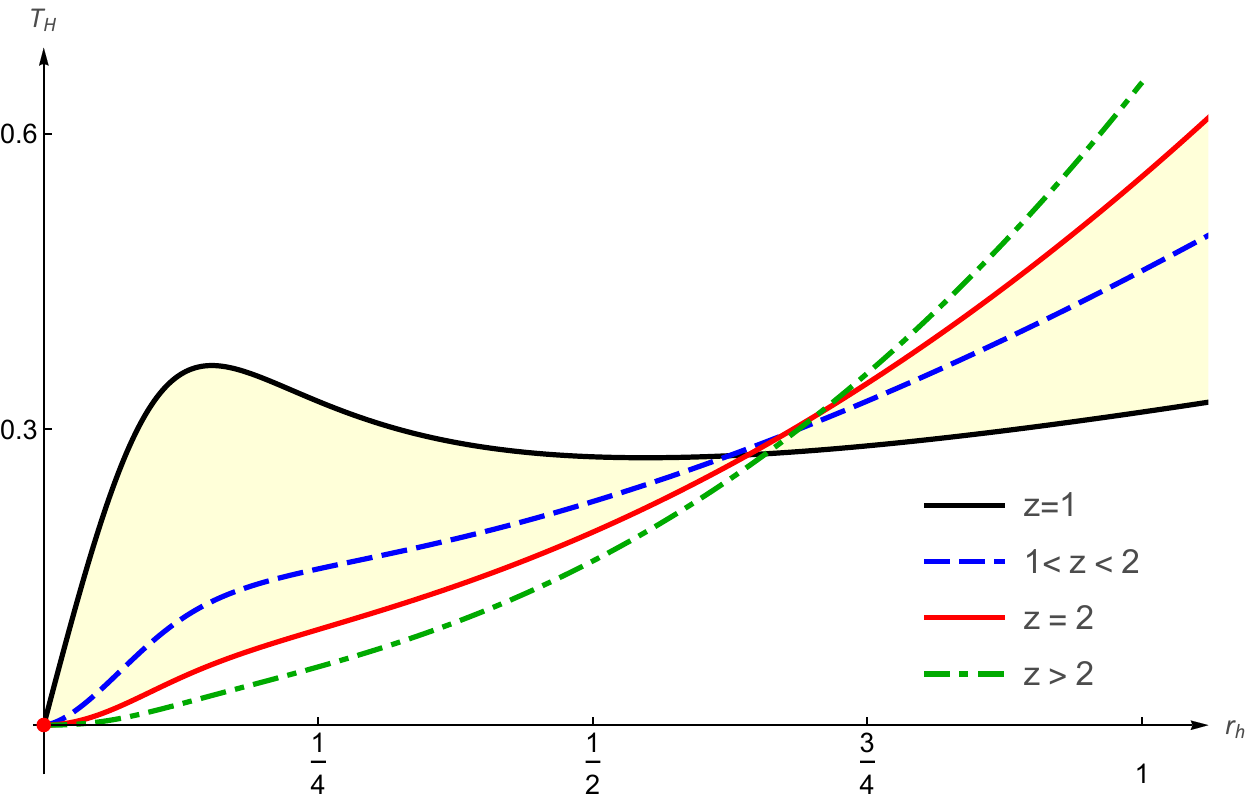}&
 \includegraphics[scale=0.2675]{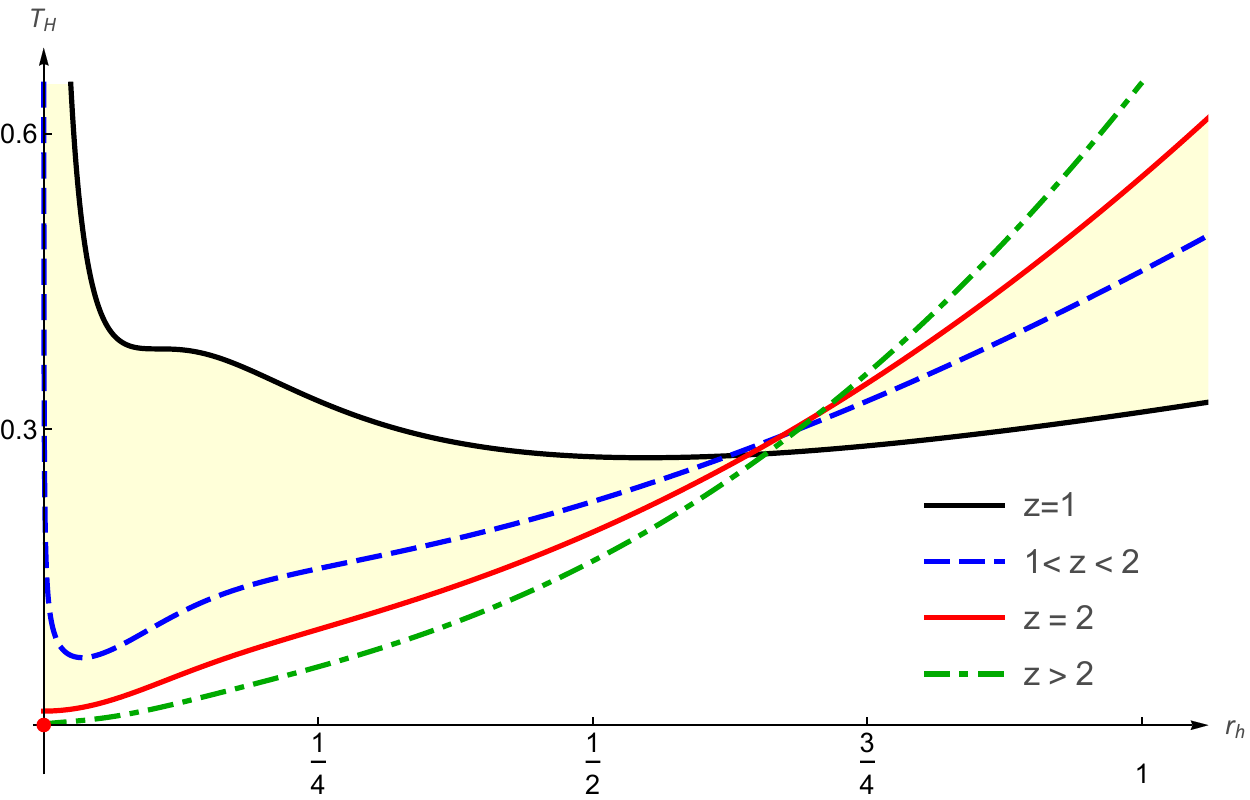}
   \\[5pt]
    \small (d) &
      \small (e)  &
      \small (f) 
  \end{tabular}
  \vspace*{8pt}
  \caption{Temperature behavior as a function of the horizon with $\gamma=1$, $\ell=1$ and $q=0.1$. In (a) we consider the case where $D>4$ with $D=5$, $\beta=22.5$ and $z=D-2+|\delta z| ~\text{(blue, dashed)},~D-2~\text{(red)},~D-2-|\delta z|~\text{(green, dotted)}~\text{and}~1~\text{(black)}$, where $\delta z=(D-1)/3$. In (b) and (c) we show for $D=4$ and different values of the $\beta q$-product, what occurs for $z<2$ and $z>2$, respectively, with $z=7/6$ and $z=12/5$. In these cases, we define $\delta \kappa=(1/2\ell-\kappa)/2$ and use $\beta q=\kappa+4\delta\kappa ~\text{(red)},~1/2\ell~\text{(yellow, dotted)},~\kappa+\delta\kappa~\text{(green, dot-dashed)},~\kappa~\text{(blue, dashed)}~\text{and}~\kappa-\delta\kappa~\text{(black)}$. In (d) and (e), and (f), we address the case where $D=4$ and $\beta q=\kappa+4\delta\kappa,~1/2\ell~\text{and}~\kappa+\delta\kappa$, respectively. In these cases, we use $z=12/5~\text{(green, dotted)}, 2~\text{(red)}, 8/5~\text{(blue, dashed)}~\text{and}~1~\text{(black)}$, respectively.   The yellow-shaded regions cover the sectors where $1\leq z<D-2$.}\label{figtemp}
\end{figure}

In this section, we study the thermodynamics of the black hole solution presented in Eq. \eqref{metricsol} from the perspective of the extended thermodynamic formalism \cite{kastor2009enthalpy,dolan2011cosmological,dolan2011pressure,kubizvnak2017black}, where pressure is defined by $P=-\Lambda/8\pi $ and the mass of the black hole is identified with its enthalpy. The associated temperature and entropy are given by
\begin{subequations}
\begin{eqnarray}
    T_{H}&=&\frac{r_h^z}{4\pi \ell^{z+1}}\left(\Delta+\frac{(D-3)\gamma \ell^2}{r_h^2}+\frac{4\beta^2\ell^2}{D-2}\left(1-\sqrt{1+\frac{q^2/\ell^2}{\beta^2}\frac{\ell^{2(D-2)}}{r_h^{2(D-2)}}}\right)\right)
,~~~~~~~\\[3pt]
    S_{bh}&=&\frac{1}{4}\left(\frac{r_h}{\ell}\right)^{D-2} \omega^{\left(\gamma\right)}_{D-2},
\end{eqnarray}\label{bht}
\end{subequations}
respectively. For finite $\beta$, in any dimension allowed by the solution \eqref{metricsol}, we have $T_H\propto r_h^z$ for large $r_h$. For small $r_h$, however, the competition between $D$ and $z$ brings relevant differences in the temperature behavior, influencing the emergence of scenarios which allow, or not, the existence of extremal regimes, identified by $T_H=0$. Indeed, the dominant contributions from the temperature as a functions of the event horizon in the region $r_h\approx 0$ are
\begin{eqnarray}\label{smalrhTemp}{\large 
  \bigg. T_{H}\bigg|_{\text{small}~r_h}\approx\left\{
\begin{array}{l}
\frac{1}{2\pi}\left(\frac{\gamma}{2\ell}- \beta q\right)\left(\frac{r_h}{\ell}\right)^{z-2}+\cdots, ~~\text{for}~~ D=4,\\[15pt]
-\frac{\beta q}{(D-2)\pi}\left(\frac{\ell}{r_h}\right)^{D-z-2}+\cdots, ~~\text{for}~~ D>4, 
\end{array}
\right.}
\end{eqnarray}
where it is possible to observe qualitative differences in temperature behavior for setups with dynamical exponent values above and below the line $z=D-2$\footnote{Although this value of the dynamical exponent is not part of the parameter set of the black hole solution found, since, in this case, the solution of the $A_a$ field is not defined.}, which acts as a wall separating distinct behaviors of $T_H$ in the non-physical sector $T_H<0$ for any $D\geq 4$. In particular, for $D>4$ there are always extremal regimes since in the limit $r_h\to 0$ the temperature diverges to minus infinity if $1\leq z<D-2$, or approaches a negative constant value in the case $z=D-2$ or forms a global minimum and returns to zero if $z> D-2$, as illustrated in Figs (\ref{figtemp}a).

For $D=4$, however, the horizon topology plays a relevant role, and its comparison with the $\beta q$-product determines the possibility of extremal solutions existing or not. In Figs. (\ref{figtemp}b) and (\ref{figtemp}c), we summarize the temperature behaviors with different values of $\beta q$ for $1\leq z<2$ and $z>2$, respectively, where we observe that for any allowed value of $z$, extremal regimes only exist if $\beta q>\gamma/(2\ell)$, indicating a lower bound on the configuration space where RN-type solutions can be found. This inequality is always satisfied in setups with planar and hyperbolic topology, but not always when dealing with spherical topology. In particular, for $r_h\to 0$ and $\gamma=1$ we observe that if $\beta q>1/(2\ell)$ one finds $T_H\to-\infty$ if $1\leq z<2$ or the temperature curve forms a global minimum and returns to zero if $z>2$, as shown in FIG. (\ref{figtemp}d), qualitatively repeating what occurs for higher dimensions. If $\beta q=1/2\ell$ we have $T_H\to 0$ as $r_h\to 0$, as illustrated in Fig. (\ref{figtemp}e). In this case, the extremal radius is zero and there are no extremal RN-type scenarios, indicating that the $\beta$-parameter acts as a suppression factor by pushing the extremal radius to zero as $\beta q \to 1/2\ell$. Finally, if $0<\beta q<1/(2\ell)$, we have S-type solutions where $T_H\to\infty$ if $1\leq z<2$ and $T_H\to 0$ for $z>2$, as shown in FIG. (\ref{figtemp}e). This last case is particularly interesting because here, for an particular interval of the configuration space defined as $\kappa<\beta q<1/2\ell$ with some $\kappa>0$ to be determined, the temperature curve exhibits up to two local minima, which indicate the emergence of a reentrant phase transition consisting of two sequential phase transitions between black hole phases, as shown in FIG (\ref{figtemp}f). The exact value of $\kappa$ can be found analytically by solving the system of equations $\partial_{r_h} T_H=0,~\partial_{r_h}^2 T_H\geq0$, where the latter inequality is saturated on the critical point, but here we choose to show how to determine it by another path, discussed in the next section.

Note also that in Figs. (\ref{figtemp}d), (\ref{figtemp}e), and (\ref{figtemp}f), for any value of the product $\beta q$, increasing the values of the dynamical exponent reduces the height of the local maximum and gives the temperature curve a monotonically increasing behavior for sufficiently large $z$. This effect results from the change which occurs in the causal structure of the solution when the marginal mass presented in Eq. \eqref{marginalmass} changes its sign from positive $(z<D/3)$ to negative $(z>D/3)$ values. The existence of such a local maximum in the temperature curve is usually associated with the presence of electric charge in the background geometry, and this reduction in height indicates that the scaling anisotropy accentuates the effect of the charge contributions in the small event horizon regime - hindering the increase in the black hole temperature as it radiates -  and acts to stabilize the system as $r_h\to 0 $, eliminating the possibility of any phase transition in the associated critical behavior. In particular, for $\beta q\geq 1/(2\ell)$, we have RN-type solutions and this elimination of the possibility of phase transitions (usually Van der Waals type) coincides with the extinction of the local maximum, in the region $z\geq D/3$. For $\beta q<1/(2\ell)$, on the other hand, we have only S-type solutions and the monotonically increasing temperature behavior only arises for $z\geq D-2$, since that although the local maximum decreases in height and eventually ceases to exist as $z$ increases (also eliminating the reentrant phase transition), the minimum associated with the Born-Infeld nonlinearity persists as long as $1\leq z<D-2$, and in these cases at least one Hawking-Page transition occurs, associated with the nonlinear dynamics of the Maxwell field.

In addition, the first law of thermodynamics in this scenario becomes \cite{gunasekaran2012extended,zou2014critical}
\begin{eqnarray}
\label{firstlaw1}dM&=&T_{H}dS_{bh}+V_{bh}dP+\Phi dQ+\mathcal{B} d\beta,
\end{eqnarray}
where $\mathcal{B}$ is the Born-Infeld vacuum polarization \cite{gunasekaran2012extended}, $V_{bh}$ represents the thermodynamic volume, $Q$ is the electric charge and $\Phi$ denotes the electrostatic potential on the
horizon. For $D>3$, the first law \eqref{firstlaw1} can be used to deduce the thermodynamical mass of the solution, given by 
\begin{subequations}
    \begin{eqnarray}
\label{eqm1}M&=&\int_0^{r_h} \Big.T_{H}\Big|_{\left(P,Q,\mathcal{B}\right)}dS_{bh},\\[3pt]
 \nonumber \label{eqm3}&=&\frac{\omega_{D-2}^{(\gamma)}}{16\pi\ell}\left(\frac{\hat{\gamma}}{\!D+z-4}\left(\frac{r_h}{\ell}\right)^{\!D+z-4}+\frac{16\pi\ell^2 P}{\!D+z-2}\left(\frac{r_h}{\ell}\right)^{\!D+z-2}+\frac{4\beta^2\ell^2}{\!D+z-2}\left(\frac{r_h}{\ell}\right)^{\!D+z-2}\left(1-\sqrt{1+\frac{q^2/\ell^2}{\beta^2}\frac{\ell^{2(D-2)}}{r_h^{2(D-2)}}}\right) +\right.\\[7pt]
 &~&~~~~~~~~\left.+\frac{4(D-2)q^2}{(D+z-2)(D-z-2)}\left(\frac{\ell}{r_h}\right)^{\!D-z-2}~{}_2F_1\left(\frac{1}{2},\frac{D-z-2}{2(D-2)};\frac{3D-z-6}{2(D-2)};-\frac{q^2/\ell^2}{\beta^2}\left(\frac{\ell}{r_h}\right)^{2(D-2)}\right)\right),
 \end{eqnarray}
 \end{subequations}
understood here as the enthalpy of the black hole. With this expression in hands, we can calculate other thermodynamical quantities, such as the volume, which becomes
\begin{eqnarray}
V_{bh}=\left.\frac{\partial M}{\partial P}\right|_{S_{bh},Q,\mathcal{B}}=\frac{\ell}{D+z-2}\left(\frac{r_h}{\ell}\right)^{D+z-2}\omega_{D-2}^{(\gamma)}, ~~~~~
\end{eqnarray}
and the electrostatic potential \eqref{maxwellsol}, which is given by 
\begin{eqnarray}
\Phi=\left.\frac{\partial M}{\partial Q}\right|_{S_{bh},P,\mathcal{B}}=\frac{q}{D-z-2}\left(\frac{\ell}{r_h}\right)^{D-z-2}\,_2F_1\left(\frac{1}{2},\frac{D-z-2}{2(D-2)};\frac{3D-z-6}{2(D-2)};-\frac{q^2/\ell^2}{\beta^2}\left(\frac{\ell}{r_h}\right)^{2(D-2)}\right),
\end{eqnarray}
which exactly sets $A(r_h)=0$ in Eq. \eqref{maxwellsol} and works as a checking condition for the internal consistency of the  expressions found so far. The Born-Infeld vacuum polarization \cite{gunasekaran2012extended} is given by
\begin{eqnarray}
\nonumber \mathcal{B}&=&\left.\frac{\partial M}{\partial \beta}\right|_{S_{bh},P,Q}= \frac{\omega^{(\gamma)}_{D-2}\beta\ell}{2(D+z-2) \pi}\left(\frac{r_h}{\ell}\right)^{D+z-2}\left(1-\sqrt{1+\frac{q^2/\ell^2}{\beta^2}\frac{\ell^{2(D-2)}}{r_h^{2(D-2)}}}~+\right.\\[7pt]
&~~&~~~~~~~~~~~~~~~~~~~~~~~~+\left.\frac{1}{2}\frac{q^2/\ell^2}{\beta^2}\left(\frac{\ell}{r_h}\right)^{\!2(D-2)}~{}_2F_1\left(\frac{1}{2},\frac{D-z-2}{2(D-2)};\frac{3D-z-6}{2(D-2)};-\frac{q^2/\ell^2}{\beta^2}\left(\frac{\ell}{r_h}\right)^{2(D-2)}\right)\right).~~~~~
\end{eqnarray}
By rearranging all these quantities, one can write down a Smarr-type relation:
\begin{eqnarray}
\label{smarr}(D+z-4)M=(D-2)T_H S_{bh}-2PV_{bh}+(D-3)\Phi Q-\beta\mathcal{B},~~~~~
\end{eqnarray}
which exactly reproduces the relation found in \cite{zou2014critical} in the limit $z\to 1$.

\section{Critical behavior}

The thermodynamic quantities calculated in the previous section can be used to study the critical behavior associated with the black hole solution \eqref{metricsol}. In particular, the equation of state derived from Eq. \eqref{bht} becomes
\begin{equation}\label{stateeq}
    P\left(T,\upsilon\right)=\frac{T}{\upsilon}-\frac{a}{\upsilon^{2/z}}-\frac{\beta^2}{4\pi}\left(1-\sqrt{1+\frac{8\pi b/\beta^2}{\upsilon^{2(D-2)/z}}}\right),
\end{equation}
where $T=T_H$ represents the associated temperature, $\upsilon=\frac{4\ell}{D-2}\left(\frac{r_h}{\ell}\right)^{z}$ denotes the specific volume and the interaction coefficients
\begin{eqnarray}\label{coefsteq}
    a=\frac{\hat{\gamma}}{\pi\left(4\ell\right)^2}\left(\frac{4\ell}{D-2}\right)^{2/z},~~~~~~
    b=\frac{2q^2}{\pi\left(4\ell\right)^2}\left(\frac{4\ell}{D-2}\right)^{2(D-2)/z}
\end{eqnarray}
are defined for simplicity. Note that the $a$-coefficient depends on the black hole topology and can be attractive, non-interacting, or repulsive for $\gamma=1,0,-1$, respectively. It implies that for any possible value of $\beta$, a critical behavior can only emerge in setups presenting spherical topology, and in this way we only need to consider the case for $\gamma=1$. For any $D\geq 4$ one finds $P(T,\upsilon)\to 0$ as $\upsilon\to \infty$ and in the small volume regime $(\upsilon\approx 0)$  one can show that the pressure behaves as
\begin{equation}\label{steq_smallbeta}
   \bigg. P\left(T,\upsilon\right)\bigg|_{\text{small}~\upsilon}\approx\frac{T}{\upsilon}-\frac{a}{\upsilon^{2/z}}+\frac{\beta}{4\pi}\frac{\sqrt{8\pi b}}{\upsilon^{(D-2)/z}}+\mathcal{O}(\upsilon^{(D-2)/z}). 
\end{equation}
This expression reveals that for $D>4$ we aways have $P(T,\upsilon)\to \infty$ as $\upsilon\to 0$, since in this case the small volume regime is dominated by the repulsive interaction contribution if $D-2>z>1$ or by the temperature term if $D-2<z$. Specifically for $D=4$, on the other hand, we have
\begin{equation}\label{steq_smallnu}
   \bigg. P\left(T,\upsilon\right)\bigg|_{\text{small}~\upsilon(D=4)}\approx\frac{T}{\upsilon}-\frac{1}{4\pi\ell}\left(\frac{1}{2\ell}-\beta q\right)\frac{\left(2\ell\right)^{2/z}}{\upsilon^{2/z}}+\mathcal{O}(\upsilon^{2/z}),
\end{equation}
and in this case for $\upsilon \to 0$ the pressure has distinct results depending on the values of the dynamical exponent and the $\beta q$ product. If $\beta q=1/2\ell$ or $z>2$ we have $P(T,\upsilon)\to \infty$ since in this case the temperature term becomes the dominant contribution. If $1<z<2$, we have $P(T,\upsilon)\to \infty$ for $\beta q>1/2\ell$  and $P(T,\upsilon)\to-\infty$ if $\beta q<1/2\ell$ because here the repulsive interaction prevails. In the large $\beta$ regime the critical behavior associated with Eq. \eqref{stateeq} is qualitatively similar to that of the model studied in \cite{moreira2024charged}, as expected. When calculating the pressure behavior for small values of $\beta$, however, the equation of state found reproduces the same expression as Eq. \eqref{steq_smallbeta} above. This coincidence shows that in the model discussed here, by increasing the intensity of the nonlinear corrections from electrodynamics one finds effective scenarios similar to those obtained by reducing the volume to considerably small values.

\begin{figure}[t!]
  \centering
  \begin{tabular}{ c @{\quad} c @{\quad} c }
  \includegraphics[scale=0.235]{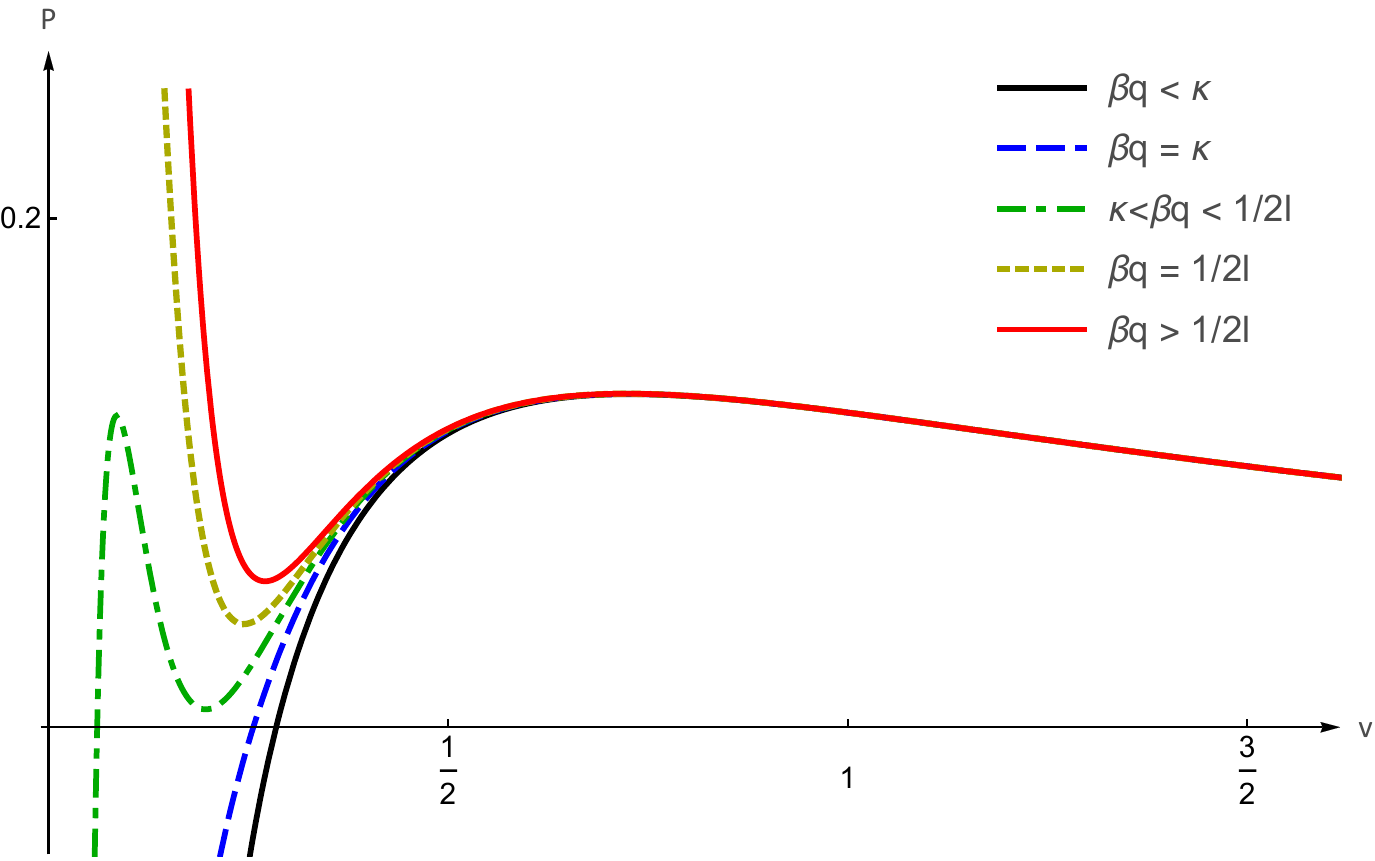} &
   \includegraphics[scale=0.235]{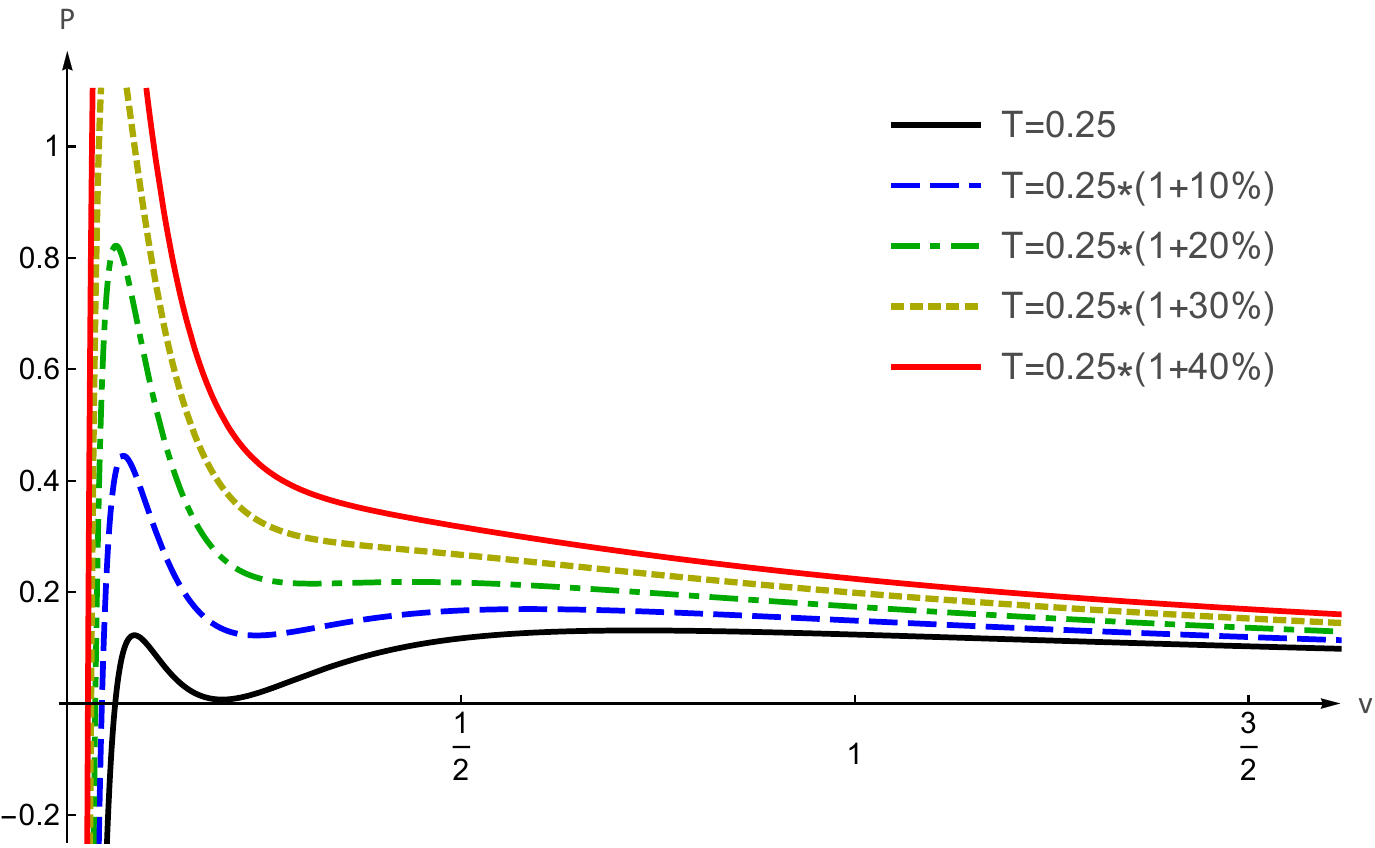}&
   \includegraphics[scale=0.235]{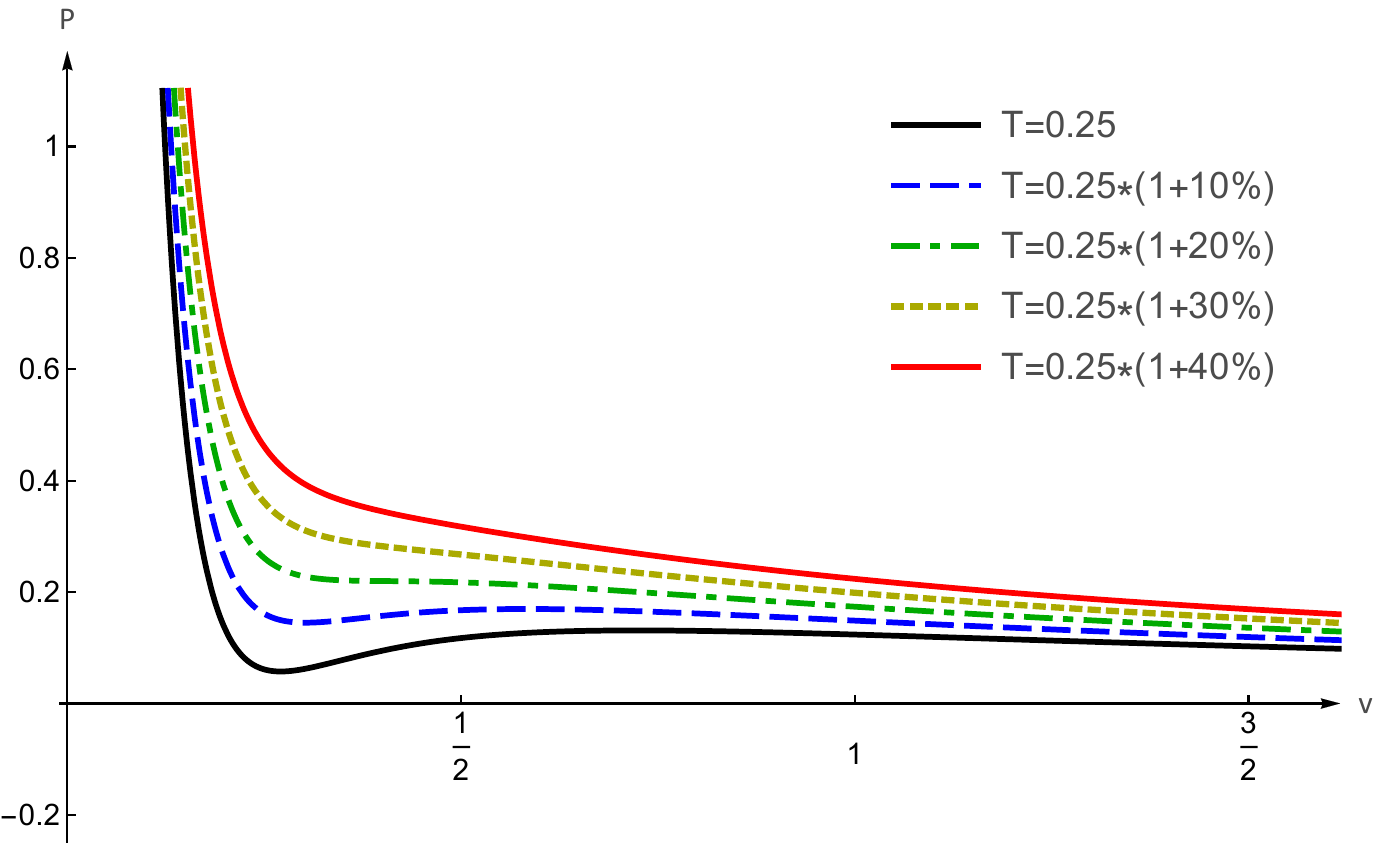}\\
    \small (a)  &
      \small (b)   &
      \small (c)  \\
    \includegraphics[scale=0.235]{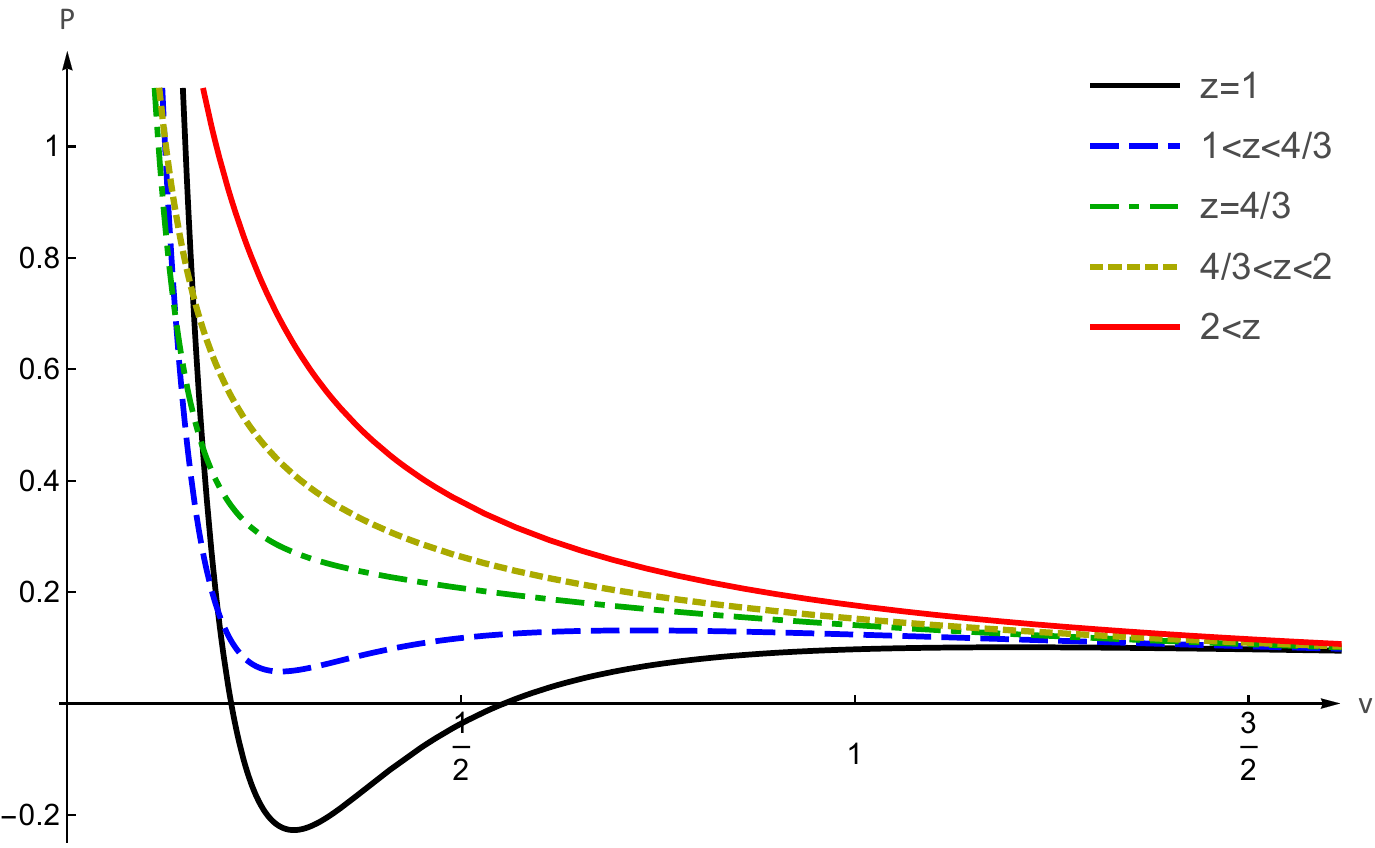} &
   \includegraphics[scale=0.235]{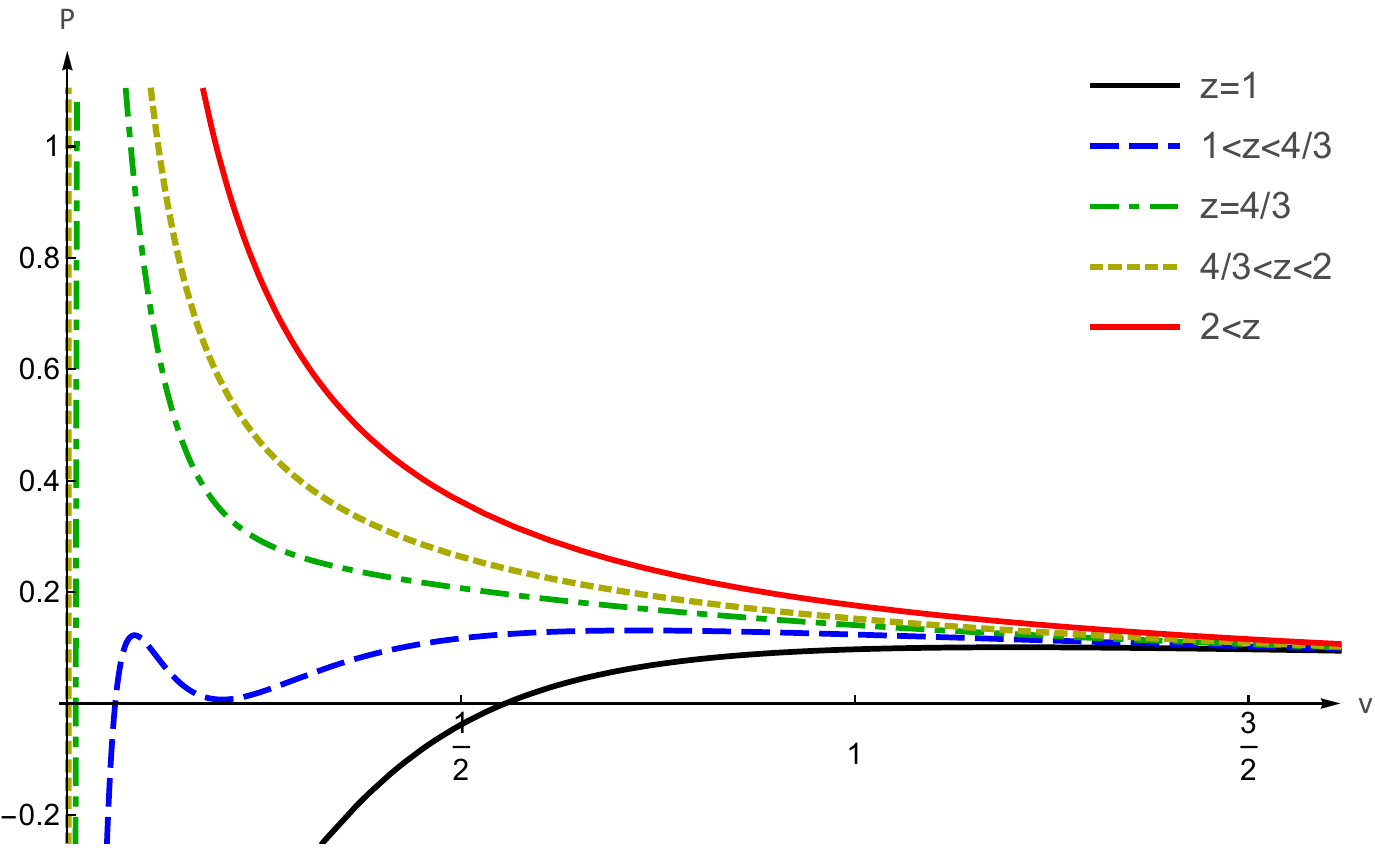}&
   \includegraphics[scale=0.235]{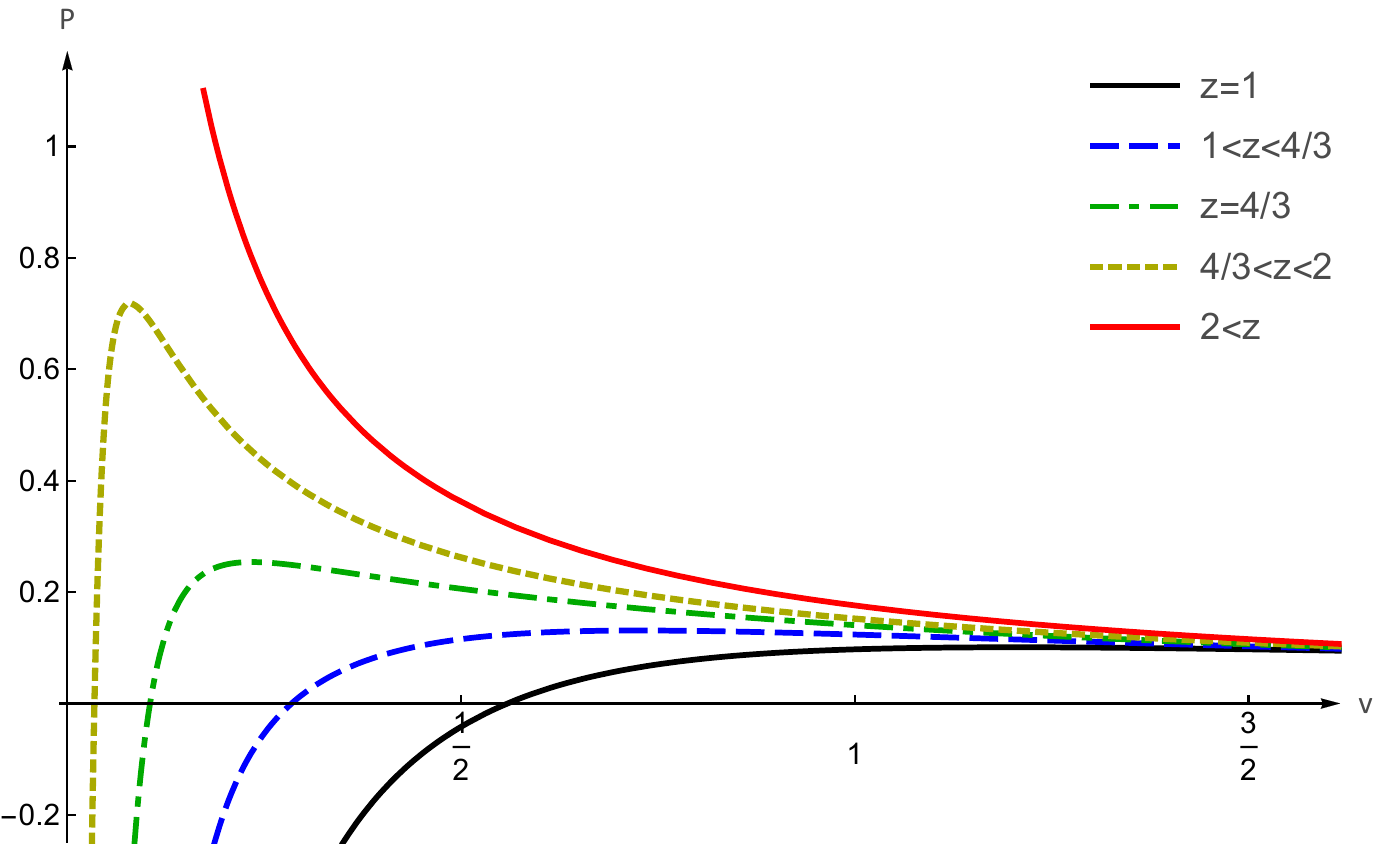}\\[5pt]
    \small (d)  &
      \small (e)  &
      \small (f) 
  \end{tabular}
  \vspace*{8pt}
  \caption{Pressure behavior as a function of specific volume with $\gamma=1$, $\ell=1$, $q=0.1$ and $D=4$. In (a), we consider the case where $\beta q$ varies and $1\leq z\leq4/3$, with $z=7/6$ and $T=0.25$. In this case, we use $\beta q=\kappa-\delta\kappa,~\kappa, ~\kappa+\delta\kappa,~1/2\ell~~\text{and}~~1/2\ell+4\delta\kappa$, with $\delta \kappa=\left(1/2\ell-\kappa\right)/2$. In (b) and (c) we illustrate how temperature influences pressure for $1\leq z\leq 4/3$ with $z=7/6$ and $\beta q=\kappa+\delta\kappa q$, for cases where $\kappa<\beta q <1/2\ell$, and $\beta q=1/2\ell+4\delta\kappa$, for cases where $\beta q \geq 1/2\ell$, respectively. In (d), (e), and (f), we consider scenarios with different values of $z$ for $\beta q\geq 1/2\ell$ (with $\beta q=1/2\ell+4\delta\kappa$), $1/2\ell>\beta q>\kappa$  (with $\beta q=\kappa+\delta\kappa$), and $\kappa\geq\beta q$  (with $\beta q=\kappa-\delta\kappa$), respectively. In these cases, we use $T=0.25$ and $z=1,~1+\delta z,~4/3,~4/3+\delta z~~\text{and}~~2+\delta z$, with $\delta z=1/6$.}\label{figpressure} 
\end{figure}

The second-order critical values of $P,~v$ and $T$ are located at the stationary inflection points of the $P-v$ diagram and can be found by solving the system of equations 
\begin{equation}
    \left.\frac{\partial P}{\partial \upsilon}\right|_{T,Q}=\left.\frac{\partial^2 P}{\partial \upsilon^2}\right|_{T,Q}=0, 
\end{equation}
which lead us to the  identities
\begin{subequations}
    \begin{eqnarray}
       T_c&=&\upsilon_c\left(\frac{2a/z}{\upsilon_c^{2/z}}-\frac{2(D-2)b/z}{\upsilon_c^{2(D-2)/z}}\frac{1}{\sqrt{1+\frac{8\pi b/\beta^2}{\upsilon_c^{2(D-2)/z}}}}\right),\\[3pt]
        P_c&=&\frac{(2-z)a/z}{\upsilon_c^{2/z}}-\frac{2(D-2)b/z}{\upsilon_c^{2(D-2)/z}}\frac{1}{\sqrt{1+\frac{8\pi b/\beta^2}{\upsilon_c^{2(D-2)/z}}}}-\frac{\beta^2}{4\pi}\left(1-\sqrt{1+\frac{8\pi b/\beta^2}{\upsilon_c^{2(D-2)/z}}}\right),
    \end{eqnarray}
\end{subequations}
for critical temperature and pressure, respectively, as a function of the critical volume $\upsilon_c$, which satisfies the equation
\begin{equation}
    \upsilon_c^{2(D-3)/z}\left(1+\frac{8\pi b/\beta^2}{\upsilon_c^{2(D-2)/z}}\right)^{3/2}=\left(1+\frac{D-z-2}{D-2}\left(1+\frac{8\pi b/\beta^2}{\upsilon_c^{2(D-2)/z}}\right)\right)\frac{(D-2)^2b}{(2-z)a}.\label{vcrit}
\end{equation}
First, note that these values exactly reproduces the critical values of the black hole solution presented in  \cite{moreira2024charged} in the limit $\beta\to\infty$ and therefore, for sufficiently high values of the $\beta$-parameter the equation \eqref{vcrit} can only have one positive real root. The number of positive real solutions of this equation tells us the number of possible second-order critical points which our system can reach but, unfortunately, we don't have a general formula to analytically capture the values of $\upsilon_c$ in any dimension. However, for $D=4$ the equation  \eqref{vcrit} can be rewritten as
\begin{equation}
    x^3-\frac{\beta^2}{8\pi b}\left(\left(\frac{4-z}{2}\right)x-\left(\frac{2-z}{4}\right)\frac{a}{b}\right)=0, ~~~\text{with}~~x=\left(\upsilon_c^{4/z}+\frac{8\pi b}{\beta^2}\right)^{-1/2},
\end{equation}
which is a reduced third-order equation with structure $x^3-k_1 x+k_2=0$ and $k_{1,2}>0$ for $1<z<2$. In this case, the analysis of the Cardano discriminant reveals that for 
\begin{equation}
    \kappa= \frac{3}{4\ell}\left(\frac{2-z}{4-z}\right)\sqrt{\frac{6}{4-z}}
\end{equation}
in the case that $\beta q\geq1/2\ell$, we have only one positive real solution, indicating that in these scenarios the system presents a single second-order critical point and behaves qualitatively similar to what occurs in typical Van der Waals phase transitions, which is expected since the associated black hole solutions are RN-type if $1\leq z<4/3$. If $1/2\ell> \beta q >\kappa$, in turn,  we have two positive real solutions, revealing that in these setups the system goes through two consecutive phase transitions and therefore exhibits a reentrant phase transition.  If $\kappa \geq\beta q$, we have no positive real solutions, which implies that in these cases the system does not exhibit a second order critical point and, in this way, one can only find a single Hawking-Page-like phase transition, associated with highly nonlinear Born-Infeld dynamics. Moreover, a direct calculation reveals that in the limit $z\to 2$ we have $\kappa\to 0$ and one of the three possible roots is shifted to infinity, while the remaining two are pushed to $\upsilon_c=0$. For $z\geq 2$, we have $\kappa<0$ (non-physical) and the system does not exhibit phase transitions.

The behavior of the $P-\upsilon$ curves associated with the equation of state \eqref{stateeq} is depicted  in Fig. \eqref{figpressure} for different scenarios with $D=4$. In Fig. (\ref{figpressure}a), we summarize the different critical behaviors found for different values of the product $\beta q$ if $1\leq z <D/3$. In these cases, we observe that the pressure can have one $(\kappa \geq\beta q)$, two $(\beta q\geq1/2\ell)$, or three $(1/2\ell> \beta q >\kappa)$ local extrema associated with Born-Infeld induced Hawking-Page-like phase transitions (one local maximum), Van der Waals phase transitions (one local maximum and one local minimum), and reentrant phase transitions (two local maxima and one local minimum), respectively. In Figs. (\ref{figpressure}b) and (\ref{figpressure}c) we illustrate the influence of the temperature for scenarios where $\kappa<\beta q<1/2\ell$ and $\beta q\geq1/2\ell$, respectively. In particular, in Fig. (\ref{figpressure}b), we observe that increasing the temperature suppresses the reentrant phase transition, but does not eliminates the Hawking-Page-like transition associated with Born-Infeld dynamics, which persists whenever $\beta q$ is sufficiently small. In Fig. (\ref{figpressure}c), on the other hand, we observe that, as the temperature increases, the system passes through a single critical point and evolves towards ideal gas-type isotherms, as is typical of Van der Waals-type systems. The dynamical exponent in the background geometry also plays a relevant role in determining which configurations allow the emergence of phase transitions. In Fig. (\ref{figpressure}d), we consider scenarios with distinct values of $z$ and $\beta q\geq 1/2\ell$, where the Van der Waals transitions associated with RN-type solutions occur and are eliminated as $z\to 4/3$. In Fig. (\ref{figpressure}e), in turn, we illustrate configurations with $\kappa< \beta q <1/2\ell$ and observe, in this case, that scale anisotropy can induce a reentrant phase transition as it moves away from the relativistic regime. Indeed, for the parameter values used in this figure, we have a scenario where if $z=1$ there is only a single Hawking-Page-like transition, but as $z$ increases a new phase transition emerges, characterizing a reentrant phase transition which is eliminated for $z \geq 4/3$. Therefore, in this setup, the presence of reentrant phase transitions can be directly associated with deviations from the relativistic regime. We can also have configurations where the system already has a reentrant phase transition in the relativistic limit and, in these cases, the increase in anisotropic scaling acts to smooth out this transition as $z \to 4/3$ and eliminate it for larger values. Finally, in Fig. (\ref{figpressure}f) we illustrate cases where $\beta q\leq\kappa$, where there is only a single Born-Infeld-induced Hawking-Page-like transition for $D=4$. In these cases, we observe that the increase in anisotropic scale values in the background geometry pushes the local maximum presented by the pressure curve upwards until it is eliminated at $z=2$, giving rise to a divergence when $\upsilon\to 0$. As expected, in none of the setups presented above one can find phase transitions if $z\geq 2$. 

\begin{figure}[t!]
  \centering
  \begin{tabular}{ c @{\quad} c @{\quad} c }
  \includegraphics[scale=0.2675]{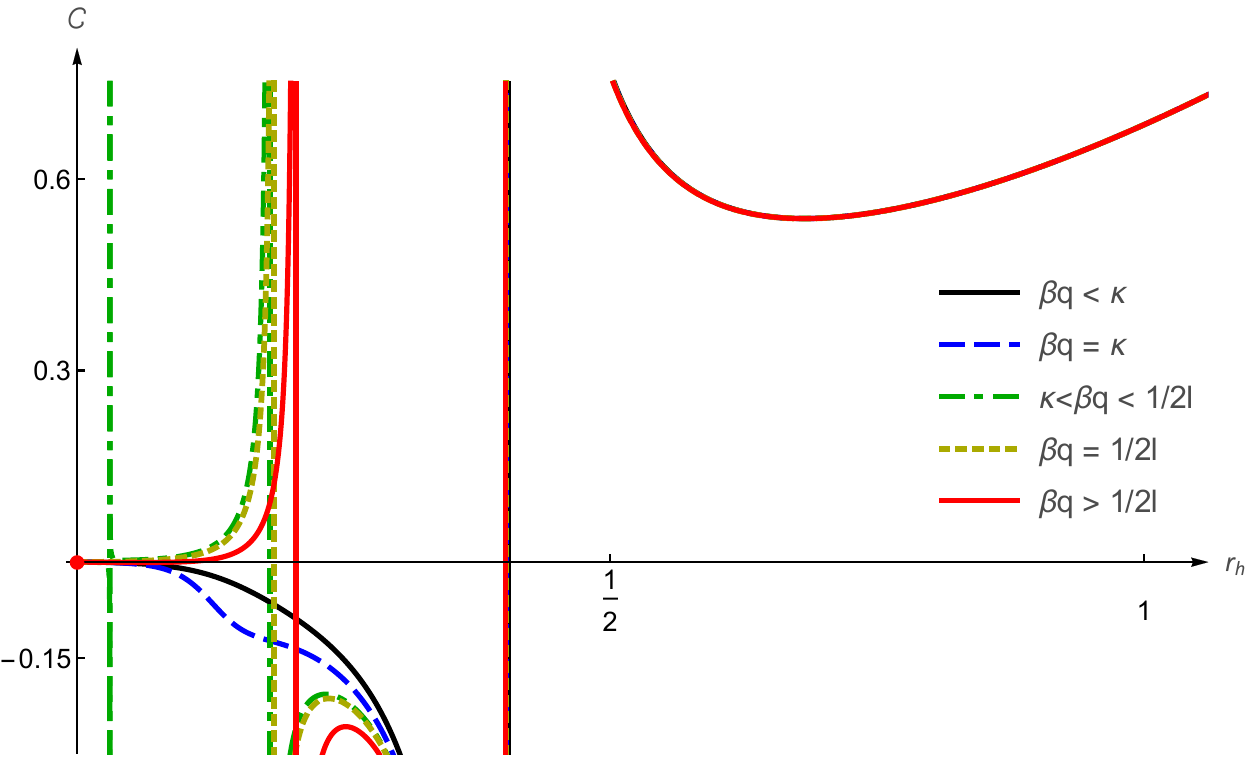} &
   \includegraphics[scale=0.2675]{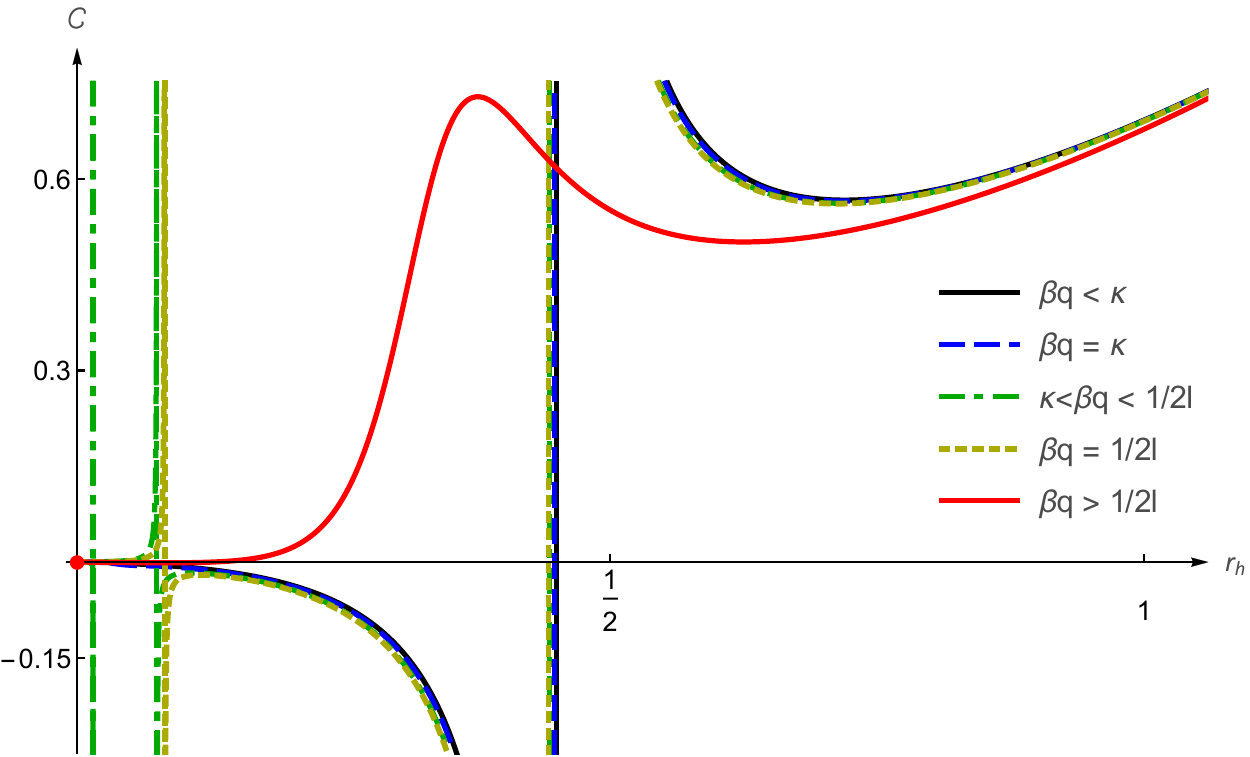}
   &
   \includegraphics[scale=0.2675]{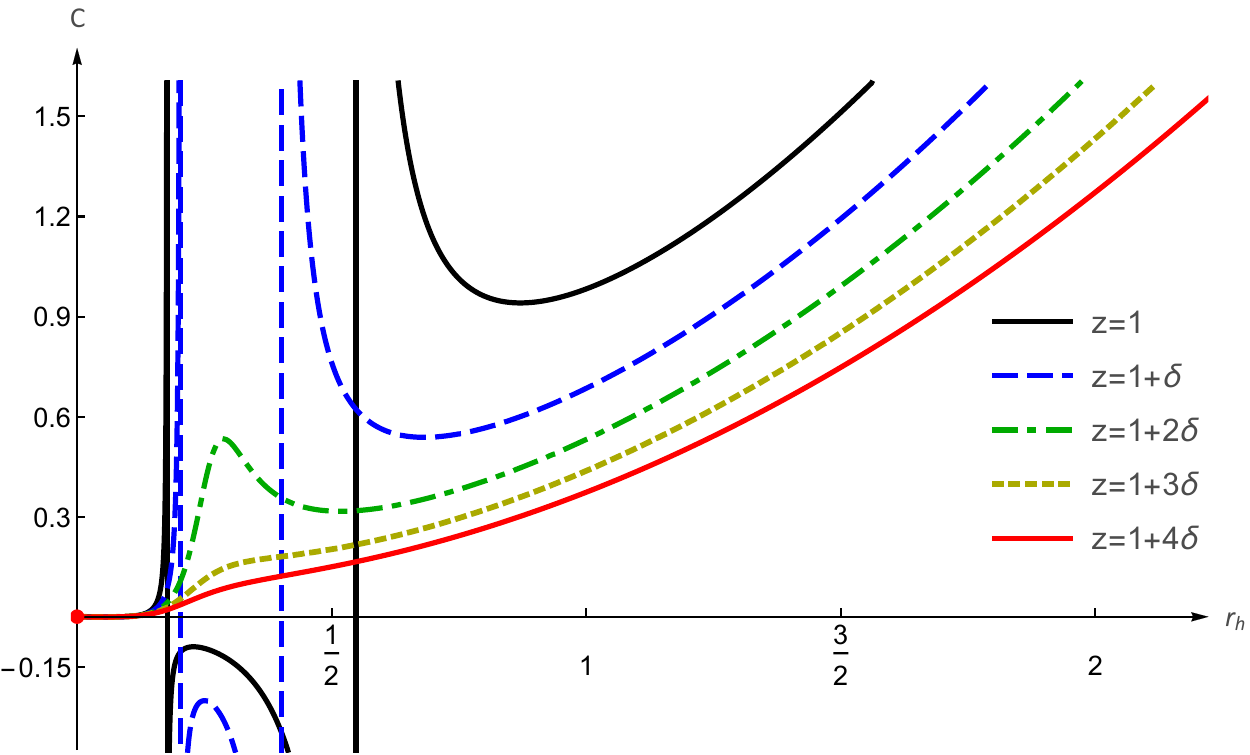}\\
    \small (a) &
      \small (b) &
      \small (c)
  \end{tabular}
  \vspace*{8pt}
  \caption{Specific heat behavior for $D=4,~\gamma=1~\text{and}~\ell=1$. In (a) and (b) we use $z=7/6$ and $\delta \kappa=(1/2\ell-\kappa)/2$, considering the cases for $\beta q=\kappa+12\delta\kappa ~\text{(red)},~1/2\ell~\text{(yellow, dotted)},~\kappa+\delta\kappa~\text{(green, dot-dashed)},~\kappa~\text{(blue, dashed)}~\text{and}~\kappa-\delta\kappa~\text{(black)}$. In particular, in (a) we fix $q=0.1$ and vary $\beta$, while in (b) we fix $\beta=10$ and vary $q$. In (c) we illustrate what happens when we move $z$ away from the relativistic limit with $q=0.1$, $\beta=10$ and $\delta =1/6$. }\label{figsh} 
\end{figure}

Another important ingredient in the analysis of the critical behavior of black holes is the specific heat, used to study the local thermodynamic stability of solutions. It can be calculated from Eqs. (\ref{bht}a,b) as
\begin{equation}\label{sh}
C = T\frac{\partial S/\partial r_h}{\partial T/\partial r_h} = \frac{\Im(r_h)}{\Re(r_h)},
\end{equation}
with
\begin{subequations}
\begin{eqnarray}
\Im(r_h) &=& 
\frac{(D-2)\omega^{(\gamma)}_{D-2}}{4z}\left(\frac{r_h}{\ell}\right)^{\!3(D\!-\!2)}\!\sqrt{1+\frac{q^2/\ell^2}{\beta^2}\frac{\ell^{2(D-2)}}{r_h^{2(D-2)}}}\!\left(\!\Delta+\frac{(D\!-\!3)\gamma\ell^2}{r_h^2}+\frac{4\beta^2\ell^2}{D-2}\left(1-\sqrt{1+\frac{q^2/\ell^2}{\beta^2}\frac{\ell^{2(D-2)}}{r_h^{2(D-2)}}}\right)\!\right)\!,~~~~\\[4pt]
  \Re(r_h) &=&\frac{4 q^2}{z}\!+\!\left(\frac{r_h}{\ell}\right)^{\!2(D\!-\!2)}\sqrt{1+\frac{q^2/\ell^2}{\beta^2}\frac{\ell^{2(D-2)}}{r_h^{2(D-2)}}} \left(\!\Delta\!+\!\left(1\!-\!\frac{2}{z}\right)\!\frac{(D\!-\!3)\gamma\ell^2}{r_h^2}+\frac{4\beta^2\ell^2}{D-2}\left(1-\sqrt{1+\frac{q^2/\ell^2}{\beta^2}\frac{\ell^{2(D-2)}}{r_h^{2(D-2)}}}\right)\!\right)\!,~~~~~~~~
\end{eqnarray}
\end{subequations}
and its behavior in $D=4$ is depicted in Fig. (\ref{figsh}) for distinct scenarios. The definition presented in Eq. \eqref{sh} implies that the specific heat must be zero at the extremal horizon and can present singularities induced by local temperature extremes. The sign reversal of the derivative which occurs near these extrema, in particular, causes sign reversals in the specific heat near these singularities, delimiting stable and unstable equilibrium regions. Due to the structures of the temperature curves for different parameter values, one can find scenarios where the specific heat exhibits one, two, or three singularities, associated with Hawking-Page-like, Van der Waals, or reentrant phase transitions, respectively, which disappear as the $z$-parameter moves away from the relativistic regime. In Figures (\ref{figsh}a) and (\ref{figsh}b), we illustrate setups with different $\beta$ and $q$ parameters, respectively, on the specific heat for $1\leq z<4/3$. In both cases, it is observed that if $\beta q \leq \kappa$, the specific heat exhibits a single divergence which separates sectors where $C<0$ and $C>0$, delimiting regions for large (stable) and small (unstable) black holes in the Hawking-Page-like transition induced by the Born-Infeld dynamics, which is consistent with the S-type behavior of the solution found in this regime of the configuration space. Furthermore, if $\kappa < \beta q < 1/2\ell$, the associated specific heat curves develop divergences which delimit four branches with alternating signs, and thus we find a typical behavior of large-small-large black hole transitions, which characterize the reentrant behavior, usual in Born-Infeld solutions with a negative cosmological constant \cite{dehyadegari2018reentrant,xu2019photon,ali2025revisiting}. For $\beta q\geq 1/2\ell$, we find scenarios whose thermodynamic structure is similar to that of RN-AdS solutions, as expected, where regions of small (stable), intermediate (unstable), and large (stable) black holes occur, characterizing a small-large black hole phase transition with Van der Waals-type behavior \cite{banerjee2012critical}. In the latter case, it should be noted that for sufficiently high values, the critical behaviors resulting from variations in the Born-Infeld parameter $\beta$ and the electric charge $q$, separately, differ significantly. For high values of $\beta$, the phase transition is always Van der Waals-type, as expected, since for $\beta\to\infty$ we recover the solution in \cite{moreira2024charged}. For sufficiently high $q$, on the other hand, it is observed that the system achieves thermal stability for all values of $r_h$.  It reveals that although the product $\beta q$ is determinant in several aspects of the solution found, the Born-Infeld dynamics, by itself, for deviations not too far from the relativistic regime (where phase transitions can occur), is not capable of providing thermal stability to the black hole solution presented here, unlike the electric charge, which can locally stabilize the system. In Fig. (\ref{figsh}c) we present scenarios for $\kappa<\beta q<1/2\ell$ with different values of the dynamical exponent and observe how the increase in scaling anisotropy eliminates phase transitions, leading to locally thermally stable scenarios for $z>4/3$ and any $r_h$. 

    \begin{figure}[t!]
    \begin{center}
\begin{tabular}{cc}
\begin{minipage}{0.5\textwidth} 
\hspace{-1cm}
\includegraphics[width=0.975\textwidth]{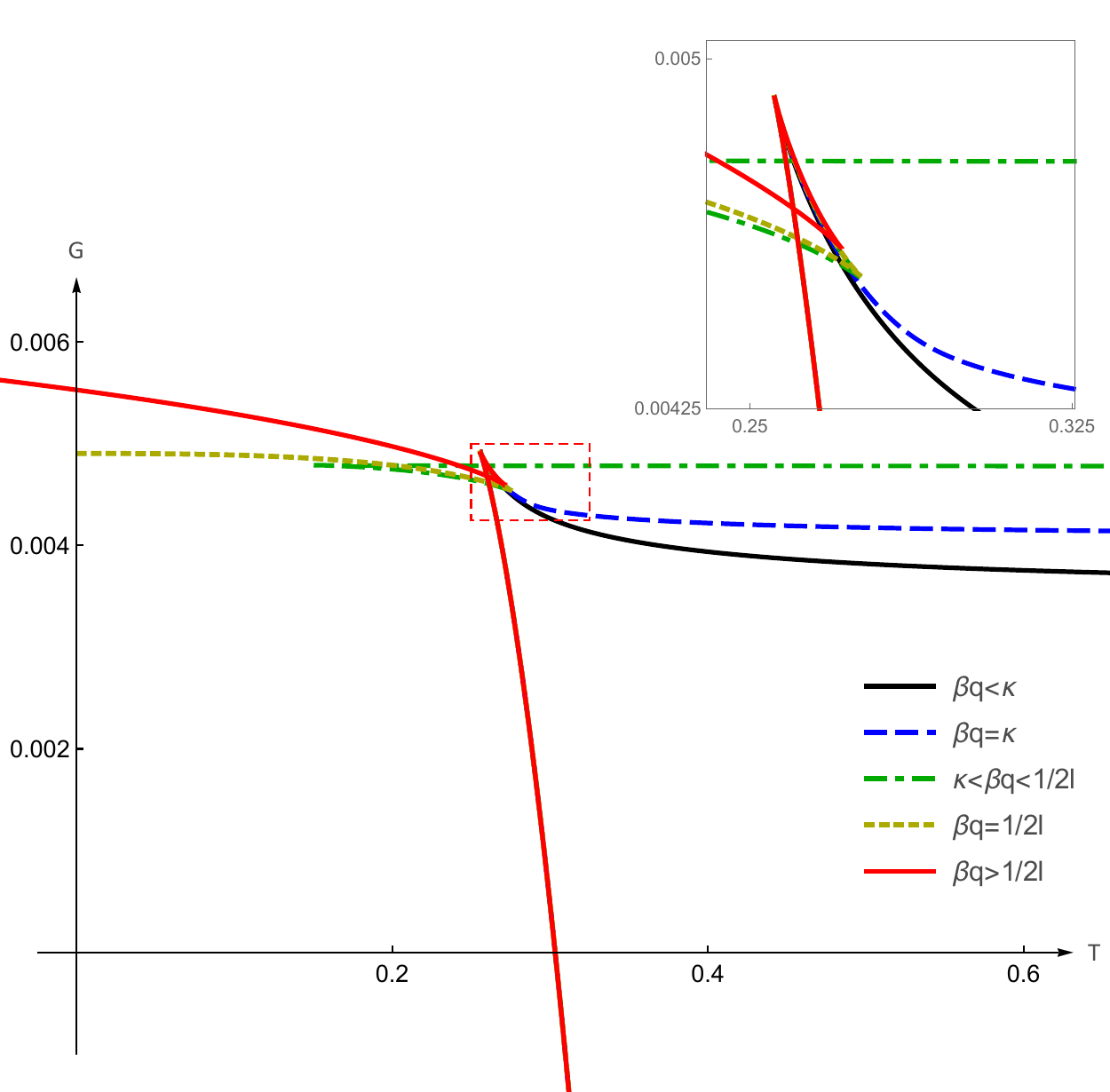}\\
\hspace{-1cm}\small (a)
\end{minipage}
\begin{minipage}{0.5\textwidth} 
\hspace{-1cm}\includegraphics[width=0.725\textwidth]{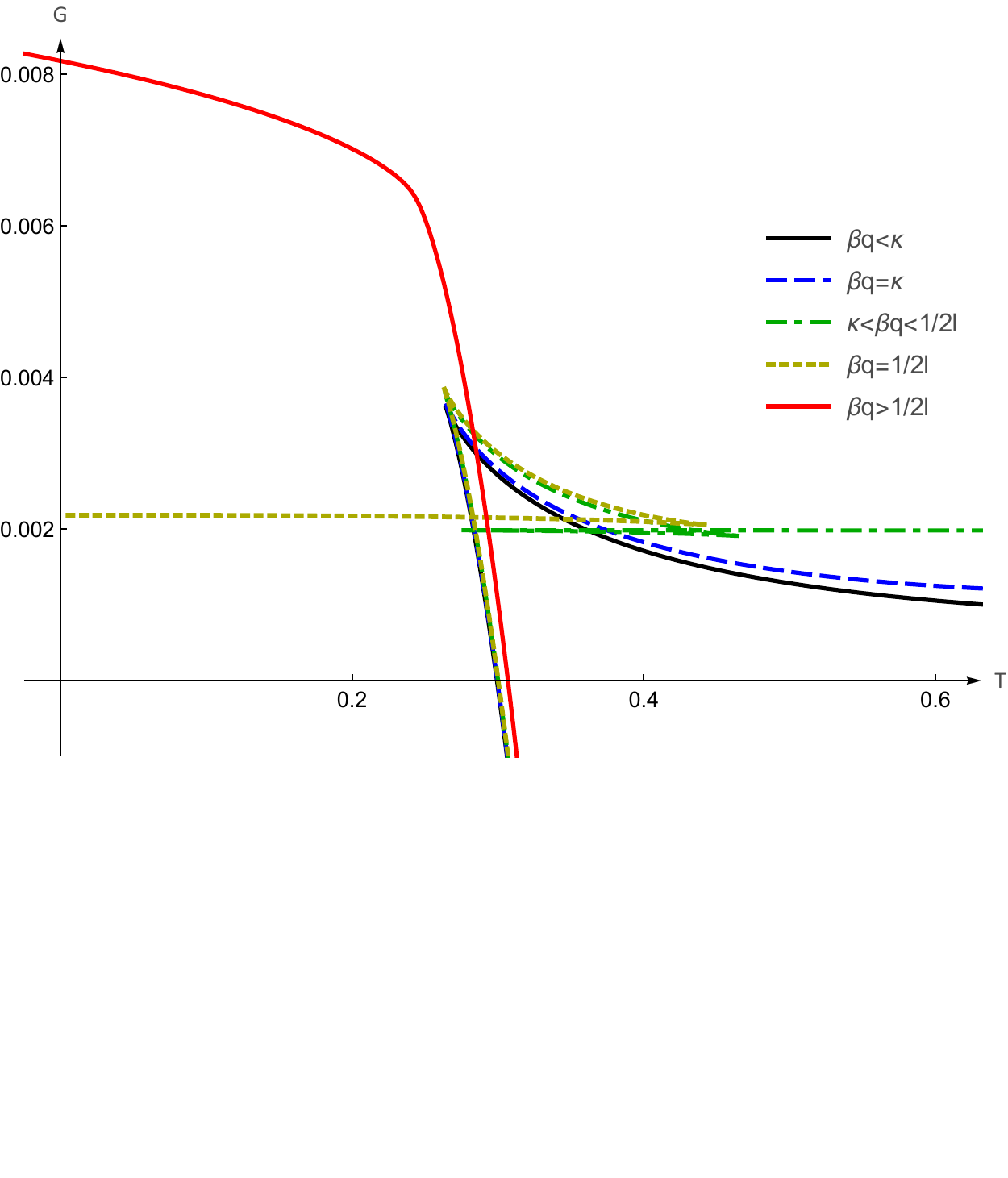}\\ \vspace{-0cm}
\hspace{-1cm}\small (b) 
\\ 
\hspace{-1cm}\includegraphics[width=0.725\textwidth]{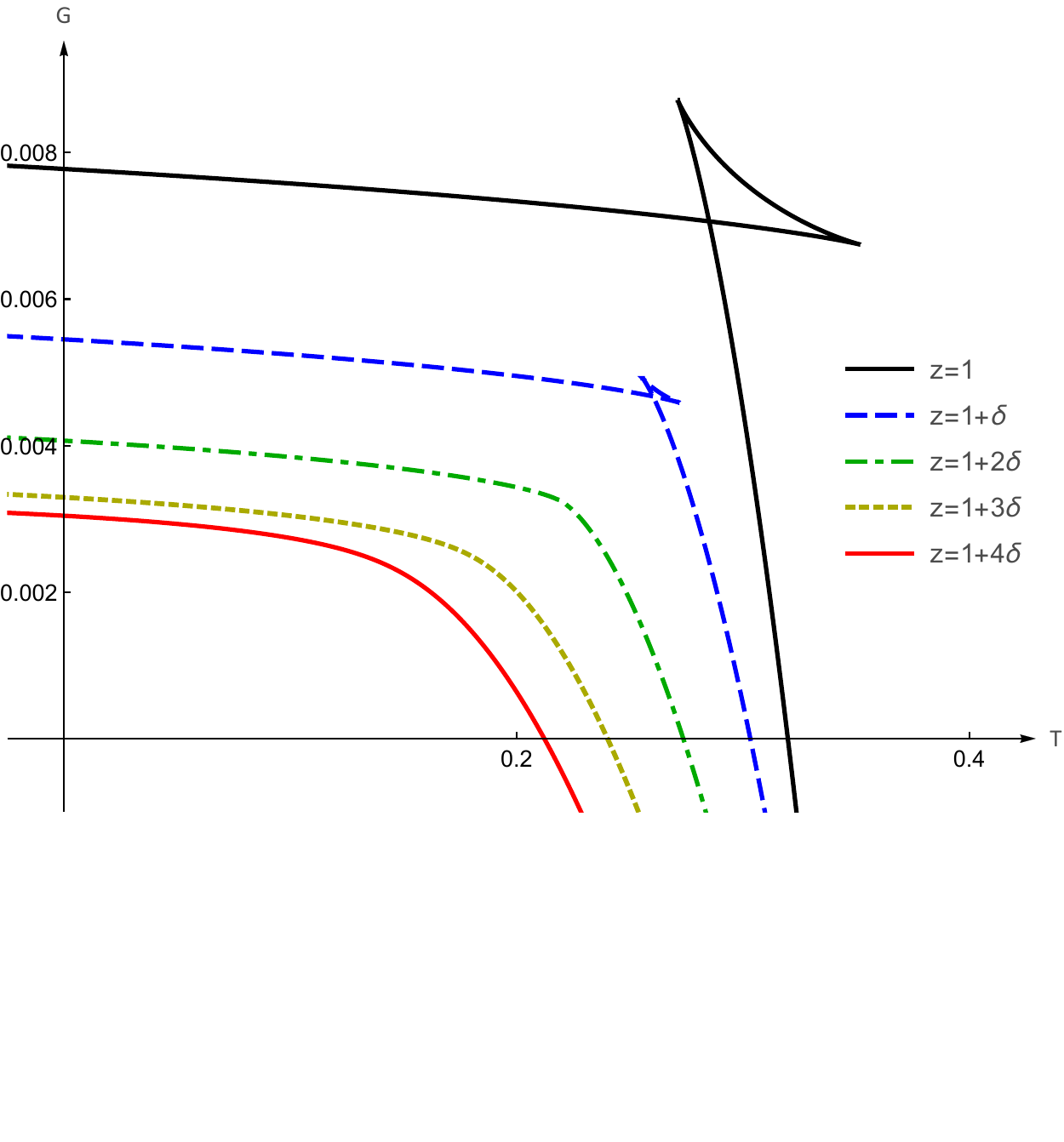}\\ \vspace{-0cm}
\hspace{-1cm}\small (c)
\end{minipage}
\end{tabular}
\caption{Gibbs free energy as a function of temperature for $D=4$, $\gamma=1$ and $\ell=1$. In (a) and (b) we use $z=7/6$ and $\delta\kappa=(1/2\ell-\kappa)/2$, considering the cases for $\beta q=\kappa+12\delta\kappa$ (red), $1/2\ell$ (yellow, dotted), $\kappa+5\delta\kappa/3$ (green, dot-dashed), $\kappa$ (blue, dashed) and $\kappa-\delta\kappa$ (black). In particular, in (a) we fix $q=0.1$ and vary $\beta$, while in (b) we fix $\beta=10$ and vary $q$. The inset in (a) shows the region where the cusps of the $G(T)$ curves occur. In (c) we illustrate what happens when we move $z$ away from the relativistic limit with $q=0.1$, $\beta=10$ and $\delta=1/6$.}\label{figfe}
    \end{center}
    \end{figure}

The global thermodynamic stability of the solution found, in turn, is addressed through the analysis of the Gibbs free energy, given by
\begin{eqnarray}\label{gibbsfe}
    G=M-TS.
\end{eqnarray}
Here, the concavity of the $G(T)$ curve delimits regions of local stability, while states with the lowest Gibbs energy determines the thermodynamically preferred configurations. In Fig. \eqref{figfe} we illustrate scenarios of $G(T)$ with $D=4$ for different values of $\beta$, $q$, and $z$. In particular, in Fig. (\ref{figfe}a) we present configurations for different values of the $\beta$ parameter, with a fixed charge $q$, where one can observe that the decrease in $\beta$-values has a negligible influence on the Hawking-Page-like temperature $(T_{0})$, which is calculated for $G=0$. This occurs because, for a fixed charge $q$ not high enough to suppress the possibility of phase transitions,  the thermodynamic effects of nonlinearity from the Born-Infeld term are only intense enough to cause significant qualitative changes in the state variables in the small $r_h$ regime (small black holes), while the crossing of the line $G=0$ occurs in the large black hole branch. This is clear, for example, in the graphic describing the temperature behavior in Fig. \eqref{figtemp}, where one can note that the temperature curves for different values of the product $\beta q$ become indistinguishable in the region immediately preceding the local minimum point associated with the presence of the negative cosmological constant. The point $G(T_0)=0$ accurs after this point and is therefore much less affected by the Born-Infeld contribution terms.  The same effect occurs for small variations in charge, as illustrated in Fig. (\ref{figfe}b), where the behavior of Gibbs free energy is presented for different values of charge $q$ and a fixed $\beta$. In a distinct behavior, the values of $r_0$ are strongly influenced by variations in the dynamical exponent $z$, as one can observe in Fig. (\ref{figfe}c). In this case, it becomes clear - now by the $G(T)$ curve - how the departure from the relativistic regime annihilates any type of transitions occurring in the system by making the formation of distinct branches impossible. A direct calculation from Eqs. \eqref{bht}, \eqref{eqm1} and \eqref{gibbsfe} performed for $\beta\to\infty$ reveals that the $T_0$ values must be concentrated in the vicinity of
\begin{equation}\label{THP}
T'_{0}=\frac{1}{\pi\ell}\left(\frac{r'_{0}}{\ell}\right)^{z}\left(\frac{3z}{4-z^2}-\frac{q^2}{(2-z)^2}\left(\frac{\ell}{r'_{0}}\right)^4\right),
\end{equation}
with associated radius 
\begin{equation}\label{rHP}
\left(r'_{0}\right)^{2}=\ell^2\frac{4-z^2}{6z^3}\left(1+\sqrt{1+\frac{12\,z^4\,(4-z)}{4-z^2}\frac{q^2}{(2-z)^2}}\;\right),
\end{equation}
 where for $q\to0$ we have $r'_{0}\to\ell$ and $T'_{0}\to 1/(\pi\ell)$ as $z\to 1$, as expected, since in this case one must recover the usual Hawking-Page temperature. Moreover, regions of global stability are delimited in the branches of the function $G(T)$, where each branch is associated with an interval of $r_h$ situated between two consecutive temperature extremes. Each local temperature extreme, in turn, generates a cusp in the $G(T)$ curve, also associated with the points where the specific heat diverges. In particular, branches with positive (negative) concavity correspond to $C>0$ ($C<0$). For a given configuration, the thermodynamically preferred phase (globally stable) is the one with the lowest Gibbs free energy, and phase transitions occur whenever there is a change from the global minimum, jumping from one branch to another. For $\beta q\leq\kappa$, as can be observed in Figs. (\ref{figfe}a,b), the curve $G(T)$ presents a single cusp associated with the only local minimum existing in the temperature curve for this setup. This cusp separates a small (unstable) branch from a large (stable) branch, and no black hole phase transition occurs, only the Born-Infeld induced Hawking-Page-like behavior. For $\beta q\geq 1/2\ell$, in turn, the curve $G(T)$ exhibits a swallowtail structure, where the small and large stable branches intersect at some temperature $T_{\ast}$, which characterizes a first-order small-to-large black hole phase transition of the Van der Waals type.

Reentrant behavior occurring in scenarios where $\kappa<\beta q<1/2\ell$ deserves additional attention because, although this inequality is a necessary condition for the emergence of the reentrant window in the state variable space, it is not sufficient. In fact, in these setups, the temperature curve exhibits three local extrema at $r_S<r_M<r_L$, and the associated temperatures $T_S$ (Born-Infeld induced local minimum), $T_M$ (charge induced local maximum) and $T_L$ (local minimun associated with the negative cosmolgical constant) induce three cusps in $G(T)$ curve. The small black hole stable branch, in particular, can only exist for $T_S\leq T\leq T_M$, while the large black hole stable branch exists for all $T\geq T_L$. In this way, the large-small-large black hole transition only occurs if the system also satisfies the condition
\begin{equation}\label{tempineq}
T_L<T_S<T_{\ast}<T_M.
\end{equation}
When the above inequality is strictly satisfied, the reentrant behavior of the phase transition can be observed as follows. As the temperature drops, the system undergoes a first-order transition from large to small black holes at $T_{\ast}$, and although the small black hole branch ends at $T_S$, the large black hole branch still persists until $T_L < T_S$, so the system returns to it via a finite jump in Gibbs free energy, which characterizes a zero-order phase transition. If the inequality in Eq. \eqref{tempineq} is violated, the existence of reentrant behavior is forbidden. Indeed, if $T_S < T_L$, the small black hole branch extends below the initial temperature of the large black hole branch, and therefore only the first-order transition occurs in $T_{\ast}$. Consequently, the zero-order transition cannot occur in this case. Furthermore, even if we have the correct ordering $T_L < T_S$, the gap between $T_S$ and $T_L$ cannot be high enough to prevent the small black hole branch from crossing the large black hole branch. Otherwise $T_{\ast}$ does not exist and thus the first-order transition does not occur, since the small black hole branch does not become a global minimum of $G(T)$. Thus, observing Figs. (\ref{figfe}a) and (\ref{figfe}b), we infer that the first one does not exhibit reentrant behavior, while the second, in turn, does exhibit it for the chosen parameter values. These differences occur because, although the product $\beta q$ plays a fundamental role in describing the behavior of the thermodynamic quantities seen so far, there are also individual contributions from $\beta$ and $q$ that affect the behavior of the thermodynamic variables and the formation of the reentrancy window. One of these differences occurs, for example, when we consider increasing the product $\beta q$ with fixed $q$ ($\beta$) and taking increasingly larger values of $\beta$ ($q$), as we did in the case of specific heat. In Fig. (\ref{figfe}a), we observe that when we increase the values of $\beta$ with fixed $q$, we return to Van der Waals-like scenarios, as expected because in this case we recover the usual dynamics of the electromagnetic field. In Fig. (\ref{figfe}b), on the other hand, we observe that by fixing $\beta$ and increasing the values of $q$, the system ends up reaching global thermal equilibrium for all temperatures, eliminating the small-black hole branches, as expected.

\section{Ending comments} 
The diffeomorphism invariance breaking has proven to be a powerful tool for finding new effective gravitational solutions. Besides providing minimal energy scalar field solutions in scenarios with probe fields (see, for instance, \cite{morris2021radially,morris2022bps,moreira2022analytical,moreira2022erratum,moreira2023localized,bazeia2025radially,luchini2026universal,andrade2026topological} and references therein) and neutral and charged Lifshitz black holes with Maxwell electrodynamics \cite{moreira2022scalar,moreira2024charged}, here we show that it is also possible to extend these ideas to setups with nonlinear electrodynamics. In particular, we focus our attention on the Born-Infeld model, due to its importance and given that the nonlinearities arising from the electrodynamic sector and the anisotropic scaling on the background geometry can induce new relevant effects, but there are other possibilities of nonlinear electrodynamics that can still be investigated in subsequent works. The qualitative behavior of the solution found depends essentially on two free parameters ($\beta$ and $q$) and we observe that a relevant part of the effects already known in the literature of Born-Infeld-AdS black holes, such as marginal mass and reentrant phase transitions, also occur in effective Lifshitz-Born-Infeld black holes if the departure from the relativistic regime is not too intense. The dependence of the marginal mass on the parameters $z$ and $\beta$ plays a fundamental role on the causal analysis of the solution, since it delimits regions with distinct causal structures and, for $m=m_0$, induces different behaviors for $D=4$, $D=5$ and $D\geq6$, but in a different way than previously described in Born-Infeld-AdS solutions due to the influence of the dynamical exponent on the background geometry. 

The thermodynamic behavior associated with the solution found is similar to that of Born-Infeld-AdS solutions in cases where the marginal mass is positive ($1 \leq z < D/3$). For $D \geq 4$, it means that Hawking-Page-like transitions can be observed, qualitatively similar to what occurs in charged AdS or Lifshitz solutions in linear electrodynamics. The case for $D = 4$ presents significant differences which are strongly influenced by scaling anisotropy. In this case, one can find Cauchy horizons only for $\beta q > 1/2\ell$, and in this setup, if $1 \leq z < 4/3$, there is only one critical point occurring in the $P-\upsilon$ curves, characterizing a Van der Waals-type phase transition which can be eliminated by sufficiently large electric charges or if $z > 4/3$. For $1/2\ell>\beta q>\kappa$, the existing solutions are S-type, and an additional critical point emerges, associated with Born-Infeld dynamics. This structure characterizes a reentrant phase transition with two critical points which respond to scaling anisotropy on the background geometry in distinct ways. The first one is associated with the presence of a charge and disappears for $z>4/3$ (similarly to what occurs in the previous case of the product $\beta q$), decharacterizing the large-small-large black hole transition which originates the reentrant phase transitions presented in the isotherms. The second is associated with the intensity of the Born-Infeld nonlinearity and is more resistant, being eliminated only for $z>2$. Thus, in this scenario, we observe that for $1\leq z<4/3$ we have a reentrant phase transition, and for $4/3<z<2$ we find a single Hawking-Page-like transition induced by Born-Infeld dynamics. If $\kappa\geq \beta q$, then there is no phase transition, but only a Hawking-Page-like transition induced purely by the Born-Infeld parameter for $1\leq z<2$. For $z>2$, all transitions are eliminated for any $\beta q$-values, leaving only scenarios with locally thermally stable solutions for any value of the event horizon. An analysis of specific heat and Gibbs free energy confirmed the nature of the transitions involved and the regions of local and global stability and instability.

Studies on gravity solutions using nonlinear electrodynamics on spacetimes with scaling anisotropy can provide a number of novel insights associated with the high degree of nonlinearity of the field equations involved, capturing effects imperceptible in linear and relativistic models and thus opening the possibility of discovering new physics. This study constitutes an effort to obtain new gravitational solutions and applications in this context, now in systems equipped with Born-Infeld electrodynamics and through the breaking of general covariance. In particular, we presented a novel black hole solution and addressed its thermodynamics and critical behavior, discussing the different possible transitions and in which regimes they can occur. We hope that these effective solutions will serve as a laboratory for testing ideas in various systems, from gravity and field theory to applications in condensed matter via gauge/gravity duality, as well as opening new avenues for future research. Some of these new approaches may include analyzing the linear stability of the solutions found and extending investigations into black hole thermodynamics with nonlinear electrodynamics in anisotropic scaling geometries. Furthermore, the method used here in the search for solutions can also serve as a basis for searching for new black hole solutions with different asymptotic behaviors.

\section*{Acknowledgments}
The authors would like to thank Roldão da Rocha for providing relevant comments and for indicating the references \cite{gunasekaran2012extended,zou2014critical}. D.C.M. would like to thank UEPB and the Brazilian agency CNPq for the partial financial support (Grant No. 402830/2023-7). F.A.B. acknowledges support from CNPq (Grant No. 309092/2022-1).
\bibliographystyle{unsrt} 
  \bibliography{biblio} 
\end{document}